\documentclass[11pt,a4paper,final]{article}

\usepackage[onehalfspacing]{setspace}
\usepackage{amsmath}    % AMS math style
\usepackage{amsfonts}   % Include AMS fonts
\usepackage{amssymb}    % Include AMS symbols
\usepackage{amsthm}     % For theorem-like environments
\usepackage{mathrsfs}   % Formal Script Math Symbol font
\usepackage{parskip}    % For whitespaced paragraphs
\usepackage{calc}       % For use of \widthof in def of \convas macro
\usepackage{booktabs}
\usepackage[authoryear,round,longnamesfirst]{natbib}
\usepackage{enumitem}   % For nicer spacing in itemized environments
\usepackage{graphicx}   % For \includegraphics
\usepackage{lpic}       % For latex-annotated graphics
\usepackage{color}      % For coloring text
\usepackage{mathtools}
\usepackage{subcaption}
\usepackage[hidelinks]{hyperref}
\usepackage{placeins}
\usepackage{accents}

\let\Oldsection\section
\renewcommand{\section}{\FloatBarrier\Oldsection}
\newcommand{\ie}{{\it i.e.}}
\newcommand{\cf}{{\it cf.}}
\newcommand{\eg}{{\it e.g.}}

\newcommand{\conv}[1]%
  {{\mathrel{\,\xrightarrow{\widthof{\,#1\,}}\,}}}
\newcommand{\convas}[1]%
  {{\mathrel{\,\xrightarrow{\widthof{\,#1\text{-a.s.}\,}}\,}}}
\newcommand{\convprob}[1]%
  {{\mathrel{\,\xrightarrow{\widthof{\,#1\,}}\,}}}
\newcommand{\convweak}[1]%
  {{\mathrel{\,\xrightarrow{\widthof{\,#1\text{-w.}\,}}\,}}}

\renewcommand{\qedsymbol}{$\Box$}

\newtheoremstyle{customtheorem}% name of the style to be used
  {0.5em}% measure of space to leave above the theorem. E.g.: 3pt
  {0.2em}% measure of space to leave below the theorem. E.g.: 3pt
  {\itshape}% name of font to use in the body of the theorem
  {}% measure of space to indent
  {\scshape}% name of head font
  {}% punctuation between head and body
  {1ex}% space after theorem head; " " = normal interword space
  {}% Manually specify head

\theoremstyle{customtheorem}

\newtheoremstyle{customremark}% name of the style to be used
  {0.5em}% measure of space to leave above the theorem. E.g.: 3pt
  {0.2em}% measure of space to leave below the theorem. E.g.: 3pt
  {}% name of font to use in the body of the theorem
  {}% measure of space to indent
  {\scshape}% name of head font
  {}% punctuation between head and body
  {1ex}% space after theorem head; " " = normal interword space
  {}% Manually specify head

\theoremstyle{customremark}

\newtheorem{hypothesisi}{Intraday Hypothesis}
\newtheorem{hypothesisa}{Closing Auction Hypothesis}
\begin{document}

\thispagestyle{empty}

\title{\vspace*{-11mm}
Effects of MiFID II on stock price formation\thanks{We thank an anonymous referee for many valuable comments on a first draft of this article. Furthermore we are grateful to Oleg Kozlovski and Victor Harmsen for providing help on data related issues and to Remco Peters for his insights on closing auction mechanisms.}}
\author{
  M. Derksen\thanks{Deep Blue Capital N.V., Amsterdam, Korteweg-de~Vries Institute for Mathematics, University of Amsterdam. Email: m.j.m.derksen@uva.nl}, B. Kleijn\thanks{Korteweg-de~Vries Institute for Mathematics, University of Amsterdam} and R. de Vilder\thanks{Deep Blue Capital N.V., Amsterdam, Korteweg-de~Vries Institute for Mathematics, University of Amsterdam} \\[1mm]
  }
\date{\today}
\maketitle

\begin{abstract}\noindent
 This paper examines effects of MiFID II on European stock markets. We study the effects of the new tick size regime, both intraday and in the closing auction. An increase (decrease) in tick size is associated with a decrease (increase) in intraday liquidity, but a more (less) stable market. In the closing auction an increase in tick size has a positive effect on liquidity. Moreover, we report a positive relationship between tick size and transacted volume, in particular in the closing auction. Finally, closing auction volumes increased heavily since MiFID II and price formation in closing auctions became more efficient.
\end{abstract}

%%%%%%%%%%%%%%%%%%%%%%%%%%%%%%%%%%%%%%%%%%%%%%%%%%%%%%%%%%%%%%%%%%%%%%%%%%%%%%%

\section{Introduction} \label{sec:intro}
At the outset of the year 2018, the new regulations for European financial markets, called the \emph{Markets in Financial Instruments Directive II }(MiFID II), came into force. MiFID II is the successor of the original MiFID\footnote{The original MiFID regulations applied since November 2007, aiming for better protection of investors.} regulations and is introduced in order to create fairer, safer, more efficient and more transparent markets. In this article the effects of MiFID II on trading stocks belonging to the STOXX Europe 600 index\footnote{The STOXX Europe 600 index has 600 components from 17 different European countries, it represents European small, mid and large cap stocks.} (hereafter referred to as the STOXX600), are examined.  We study the effects of the new tick size regime and also report significant effects that are not related to the tick size regulations, especially in the closing auction. 

Several parts of MiFID II are relevant to stock trading and market microstructure. An important part of the regulations is a new tick size regime.\footnote{The new tick size regime is introduced in Article 49 of MiFID II, see \url{https://www.esma.europa.eu/databases-library/interactive-single-rulebook/mifid-ii/article-49}} The \emph{tick size} refers to the smallest possible price increment when trading an asset. During the last decade, exchanges have reduced their tick sizes in order to create tighter bid-ask spreads. However, since a few years it is recognized that this has had a negative effect on the quality of the market. That is why with MiFID II a new tick size regime was introduced, setting a minimum tick size for every traded instrument on every exchange in Europe, taking into account the price and liquidity of the instrument. Other relevant parts of MiFID II include the double volume cap (DVC) and the \emph{best execution} rule. The DVC limits the volume that is allowed to be traded in the so-called \emph{dark pools} (private exchanges in which liquidity is not visible a priori), in order to protect the transparency of the market.\footnote{Only 8\% of the total transacted volume (over all trading venues) in a stock is allowed to be carried out in dark pools, and only 4\% of the total transacted volume is allowed to be carried out in a single dark pool. This is stated in Article 5 of MiFIR, see \url{https://www.esma.europa.eu/databases-library/interactive-single-rulebook/mifir/article-5}} The best execution rule mandates intermediaries to obtain the best possible price when executing a client's order.\footnote{This is stated in Article 27 of MiFID II, see \url{https://www.esma.europa.eu/databases-library/interactive-single-rulebook/mifid-ii/article-27}} A similar rule already existed in the previous MiFID regulations, but the obligation to prove fulfillment has become much more urgent, possibly changing the way orders are executed.

We first study the effects of the new tick size regime. The French regulator did a first examination of the effects of the new tick size regime and concluded the regime had positive effects on liquidity and market stability, see \cite{Amf18}. There already exists a large body of literature on the effects of changes in tick size, most of it based on the change to decimal pricing in the US in 2001 (see \eg\ \cite{Bessembinder03, Chakravartyetal04,Chakravartyetal05} and many others), but also on other moments in time and other exchanges (see \eg\ \cite{Andersonpeng14,Bacidore97,Ronenweaver01}).  See \cite{Verousisetal18} for a very recent and exhaustive review of the literature on tick sizes. Most of the existing literature on this subject is relatively old (before the rise of high frequency trading) and/or based on only trade- and best bid/ask-data. However, very recently a new line of studies emerged that investigate stocks that were subject to the tick size pilot in the US\footnote{In 2016 the US Securities and Exchange Commission launched the Tick Size Pilot Program as an experiment to assess the effects of an increase in tick size, for illiquid stocks mainly.} (see for example \cite{Chungetal20} and \cite{Griffithroseman19}, among others). In today's financial markets most securities are traded in continuous double auctions. However, to determine a daily closing price, a closing call auction is often used. In these closing auctions orders are collected for a while without giving rise to trades immediately, after which a closing price is determined and every possible transaction is conducted against this price.\footnote{See appendix \ref{appendix:closing auctions} for more information on closing auction mechanisms.} All of the abovementioned studies focus on the effects of tick size changes on intraday continuous trading, but none of them studies the effects of tick size on trading in the closing call auction. In this article, we fill this gap by studying the effects of the new tick size regime on both intraday continuous trading and trading in the closing auction.  An increasingly large part of the daily transacted volume is transacted in the closing auction, implying that studying the call auction gets even more important. Unfortunately, the literature on the call auction is fairly scarce compared to the enormous (both empirical and theoretical) literature on continuous trading. Recent theoretical contributions are by \cite{Toke15} and \cite{Derksenetal20}, while the empirical literature mainly focuses on the effects of call auction mechanisms on price formation  and the difference with continuous trading (see \eg\ \cite{Hillionsuominen04,Kandeletal12, Paganoschwartz03}, among others).  

 We study how liquidity, transacted volumes, volatility and market stability are affected by a change in tick size, for both intraday trading and in the closing auction. We find that the increase in depth after an increase in tick size can not offset the contemporaneous increase in bid-ask spread (except for very large orders), implying that intraday liquidity for small orders is negatively affected by an increase in tick size, while it is positively affected by a decrease in tick size (in line with for example \cite{Goldsteinkavajecz00}). In the closing auction there is no bid-ask spread, but depth still increases, leading to improved liquidity in closing auctions after an increase in tick size. Furthermore, we find that an increase (decrease) in tick size leads to a moderate increase (decrease) in intraday transacted volumes and a strong increase (decrease) in transacted volumes in the closing auction. Finally, we show that an increase (decrease) in tick size has a positive (negative) effect on the stability of the market.

Moreover, we also document significant post-MiFID II effects that can not be attributed to the tick size regime, especially in the closing auction. We find that since the introduction of MiFID II an increasing part of the daily transacted volume is transacted in the closing auction, mainly driven by an increase in limit order volume, not market order volume. Along with the increase in volumes, price deviations in the closing auction increased. 
 Pre-MiFID II a high absolute return in the closing auction was followed by a correction the next day, but post-MiFID II this mean reversion has vanished. These observations suggest that the extra volume in the closing auctions contributes to price formation and makes closing prices more efficient, ruling out the possibility that the extra volume is caused by the increase in passive investing. We find no evidence that the DVC rule has caused volume to move from dark pools to the regular exchanges (in line with \cite{Johannetal19}) and we propose the best execution rule as a possible explanation of the observed effects.
 
This paper constitutes a first large scale examination of the effects of MiFID II on European equity trading, documenting both expected and unexpected consequences of this regulatory intervention, which are of interest to academics, policy makers and market practitioners. Particular focus goes to tick size effects, especially regarding the influence of  tick sizes in the closing call auction (rather than intraday continuous trading, which has been the exclusive focus of most existing literature). In addition, we investigate the effects of tick size on market stability, a topic that has received less attention in the literature. Finally, we contribute to the literature on call auctions, by studying tick size effects and the efficiency of price formation in the closing auction.

The remainder of this article is organized as follows. In section \ref{sec:hypos} we formulate several testable hypotheses on how the new tick size regime can affect trading in the intraday continuous market and in the closing auction. In section \ref{sec:data} the data is discussed and the method is explained. In sections \ref{sec:liqudity} and \ref{sec:stability} the effects of the new tick size regime  are examined. In section \ref{sec:closing auctions} we discuss post-MiFID II effects that can not be attributed to the new tick size regime. Finally, in section \ref{sec:conclusions} the results are summarized and concluding remarks are made. In appendix \ref{appendix:closing auctions} information on different closing auction mechanisms can be found and appendix \ref{appendix:darkpools} reports results on the effects of the DVC rule.
%%%%%%%%%%%%%%%%%%%%%%%%%%%%%%%%%%%%%%%%%%%%%%%%%%%%%%%%%%%%%%%%%%%%%%%%%%%%%%%
\section{Hypotheses development} \label{sec:hypos}
In this section we will formulate testable hypotheses for the effects of the new tick size regime on intraday continuous trading and on trading in the closing auction.
\subsection{Liquidity, trading costs and transacted volume}
There exists a large empirical literature on the relationship between tick size and liquidity. Generally, the focus is on the bid-ask spread (defined as the difference between the lowest ask price and the highest bid price) and market depth (the volume that is accumulated in the order book). The literature unanimously shows that a higher (lower) minimum tick size leads to a higher (lower) bid-ask spread and higher (lower) volumes at the best bid and ask (see \eg\ \cite{Alampieskilepone09,Apgwilymetal05,Biaisetal10,Bollenwhaley98,Chanhwang02,Chungetal05,Hsiehetal08,Vannessetal00}, among others). An increase (decrease) in spread after an increase (decrease) in tick size, is generally explained by the extent to which the tick size is a binding constraint for the spread: if the spread is oftentimes only one tick, an increase in tick size will mechanically increase the spread, while a decrease in tick size will offer the opportunity to quote lower spreads. Even if the tick size is not binding the spread, market participants can place limit orders at tighter prices when the tick size is lower. Reflecting this second argument, \cite{Bessembinder00} finds that spread and tick size are also positively related when the tick size is non-binding. Furthermore, a positive relationship between depth and tick size is explained by the following arguments. Volumes at the best bid and ask will be higher when the tick size is higher, because a coarser price grid will cause volume to concentrate more on the same price level. Also, the cost of front running increases with the tick size, making it more likely a market participant decides to join the queue, instead of crossing or tightening the spread.
 Following \cite{Goldsteinkavajecz00}, we associate a decrease in spread with improved liquidity for smaller orders, while increased depth suggests improved liquidity for larger orders. We study the tradeoff between spread and depth by investigating the price impact of market orders of increasing size.
The first hypothesis thus is as follows.
\begin{hypothesisi} \label{hyp:intraday_liq}
After an increase (decrease) in tick size, the bid-ask spread and market depth will increase (decrease). Consequently, the trading costs for small trades increase (decrease) with an increase (decrease) in tick size, while trading costs for large trades decrease (increase).
\end{hypothesisi}
In the closing call auction there is no bid-ask spread, because bid and ask quotes overlap during the accumulation phase of the auction. However, the limit order book just after the clearing of the auction can be interpreted in a similar way. The post auction spread (defined as the bid-ask spread observed just after clearing) measures liquidity around the closing price: when this spread is large, the range of possible clearing prices is large, indicating that an extra market order might have altered the closing price strongly. But even more than in continuous trading, depth determines how easily the price is moved: when volume in the post auction limit book is large, an extra market order would have zero price impact. This is an important difference with continuous trading, where crossing the spread always induces trading costs, independent of order size. Similar to the case of intraday trading, we expect tick size and depth to be positively related, because a higher tick size causes volume to concentrate more on the same price level. We hypothesize that depth is a more important determinant of auction liquidity than the post auction spread and we study the tradeoff by looking at price impact. We thus propose the following hypothesis. 
\begin{hypothesisa} \label{hyp:auction_liq}
An increase (decrease) in tick size will lead to a higher (lower) post auction spread and to higher (lower) depth. Consequently, an increase (decrease) in tick size will have a positive (negative) effect on closing auction liquidity.
\end{hypothesisa}

As we have seen, the relationship between tick size and intraday liquidity is rather complex (basically a tradeoff between spread and depth), which raises the question how this affects transacted volume.
The relationship between tick size and transacted volume (if it exists) is not well understood in the literature. \cite{NiemeyerSandas94} find (some) evidence that tick size and transacted volume are negatively related, a relationship that is also predicted by the theoretical work of \cite{Harris94} and \cite{Goettleretal05}. This negative relationship is supported by the compelling argument that lower tick sizes lead to lower spreads, causing market participants to trade more.  On the other hand, the zero intelligence model (where traders make decisions exogenously) of \cite{ChiarellaIori02} predicts that transacted volume and tick size are positively related. Indeed, if traders do not alter their decisions based on the changing tick size, a larger tick size will cause more orders to be rounded to the same price, making matching of buy and sell orders more likely. In line with this argument, \cite{Chakravartyetal04} find that transacted volumes decrease after a decrease in tick size. However, \cite{Ahnetal96, Ahnetal07}, \cite{Andersonpeng14}, \cite{Bacidore97}, \cite{Geraceetal12} and \cite{Keetal04} do not find evidence of effects of tick size reduction on transacted volumes. More recently, \cite{Oharaetal18} also find that total transacted volume is not affected by the tick size and the \cite{Amf18} finds that transacted volumes are not impacted by the new tick size regime.  Clearly, there is no consensus on the relationship between tick size and transacted volume, but most empirical evidence (and importantly, very recent evidence) suggests that there is no significant relationship between tick size and transacted volume. Based on this, we propose the following hypothesis.
\begin{hypothesisi}\label{hyp:intraday_tv}
A change in tick size will not affect the intraday transacted volume.
\end{hypothesisi}
In the closing auction, there is no bid-ask spread, ruling out the channel through which a decrease in tick size may have a positive effect on transacted volume. On the other hand, the reasoning that a larger tick size may lead to larger transacted volumes, because it will cause more orders to overlap, does still hold. This mechanical argument is even stronger in the closing auction, where less interaction between market participants is possible.  So we propose the following hypothesis.
\begin{hypothesisa}\label{hyp:auction_tv}
An increase (decrease) in tick size will have a positive (negative) effect on transacted volume in the closing auction.
\end{hypothesisa}
\subsection{Volatility and market stability}
Most of the relevant empirical literature suggests that a higher (lower) tick size leads to higher (lower) volatility (see \eg\ \cite{Hau06,Keetal04,Ronenweaver01}). This is also in line with Intraday Hypothesis \ref{hyp:intraday_liq} on bid-ask spreads, as higher spreads are associated with higher volatility (see \eg\ \cite{Keetal04}). Furthermore, a positive relationship between tick size and volatility is also supported by theoretical results of \cite{ChiarellaIori02}, who argue that an increase in minimal price change will lead to a linear increase in price deviations. So our hypothesis concerning volatility reads as follows.
\begin{hypothesisi} \label{hyp:intraday_vola}
Volatility will increase (decrease) after an increase (decrease) in tick size.
\end{hypothesisi}
Concerning the closing auction, we are interested in the absolute deviation of the closing price with respect to the price just before the auction, as a measure of the auction's volatility. Again, we expect that a higher tick size will lead to larger price deviations, because the minimal price change is larger. Therefore, we formulate the following hypothesis.
\begin{hypothesisa} \label{hyp:auction_vola}
Absolute auction returns will increase (decrease) after an increase (decrease) in tick size.
\end{hypothesisa}
Finally, we will adress the issue of market stability. One of the aims of the new tick size regime was to create a more stable market, as it was noted that too small tick sizes lead to unstable markets \citep{Laruelleetal18}. To reduce microstructural noise, a lower number of bid-ask price changes and a lower number of (small) trades, combined with a higher trade size are desired and this should be achieved by an increase in tick size. A larger tick size will increase the cost of front running and the cost of crossing the spread, which leads to less microstructural noise (frequent small price movements). We therefore formulate the following hypothesis.
\begin{hypothesisi}\label{hyp:intraday_stability}
An increase (decrease) in tick size will have a positive (negative) effect on the stability of the market: an increase (decrease) in tick size will lead to a decrease (increase) in the number of best bid-ask price updates and the number of (small) trades.
\end{hypothesisi}
In the closing auction less microstructural noise would mean that the indication price is updated less often during the auction, reflecting a more stable price formation.\footnote{During the closing auction, indication price and volume are released continuously, indicating the price and volume if the auction would be cleared at that time. See appendix \ref{appendix:closing auctions} for more information on auction mechanisms and rules.} The corresponding hypothesis is as follows.
\begin{hypothesisa}\label{hyp:auction_stability}
An increase (decrease) in tick size will be followed by a decrease (increase) in the number of indication price updates, reflecting a more stable price formation process.
\end{hypothesisa}
\section{Data and method}
\label{sec:data}
Two years (2017 and 2018, before and after the new regulations) of intraday order book data (minimal 5 levels deep on both sides) and transaction data is obtained for all constituents of the STOXX600 index. The STOXX600 consists of 600 components, representing 17 countries in Europe\footnote{Those are: Austria, Belgium, Denmark, Finland, France, Germany, Ireland, Italy, Luxembourg, the Netherlands, Norway, Poland, Portugal, Spain, Sweden, Switzerland and the United Kingdom. Around 75\% of the index is listed on Euronext (Amsterdam, Paris, Brussels, London, Oslo, Dublin, Lisbon), XETRA, SIX or Borsa Italiana.}. Some stocks in the STOXX600 are traded on several exchanges and in different currencies, we only consider trading on their European primary exchanges\footnote{Some companies have multiple European primary exchanges (\eg\ Unilever is listed in both Amsterdam and London), in that case we choose the one with the highest trading volume (on average).}. The stocks are divided into three groups, based on the market capitalization of the stock, which leads to 200 small cap stocks, 200 mid cap stocks and 200 large cap stocks. Stocks that were not in the STOXX600 index for one of the years are excluded. 

To examine the effects of the new tick size regime, the stocks are divided into groups based on the way their tick size is affected by this new regime: a group of which the tick size increased (ts$\uparrow$), a group of which the tick size decreased (ts$\downarrow$) and a group of which the tick size remained unchanged (ts$\leftrightarrow$). The last group acts as a control group for the other two groups. Because tick sizes can (as well in the old as in the new tick size regime) also change during a calendar year due to changes in the stock price, the groups have the following definitions. 
\begin{itemize}
\item {\bf ts$\uparrow$}: every stock whose average tick size over 2018 is a factor\footnote{All the results discussed in this article are robust with respect to a reasonable choice of the factor 1.5.} 1.5 higher than its average tick size over 2017. This group contains 179 stocks, of which 64 small caps, 68 mid caps and 47 large caps. 
\item {\bf ts$\downarrow$}: every stock whose average tick size over 2017 is a factor 1.5 higher than its average tick size over 2018.
This group contains 144 stocks, of which 42 small caps, 43 mid caps and 59 large caps.
\item {\bf ts$\leftrightarrow$}: all other stocks.
This group contains 220 stocks, of which 80 small caps, 70 mid caps and 70 large caps. 
\end{itemize}
In table \ref{table:stats} the average tick size change and the average market capitalization of the groups is shown.
\begin{table}[h]
\small
\begin{center}
\begin{tabular}{ p{0.8cm} p{1.9cm} p{1.9cm} p{1.9cm} p{1.9cm} p{1.9cm} p{1.9cm}}
 \hline
&\multicolumn{2}{c}{Small cap stocks} & \multicolumn{2}{c}{Mid cap stocks} &\multicolumn{2}{c}{Large cap stocks} \\
& $\Delta$tick size & market cap & $\Delta$tick size & market cap & $\Delta$tick size& market cap\\
ts$\uparrow$ & $+132.2$\%  &3.85 &+111.8\% & 8.94 & +137.3\% & 42.1 \\
ts$\downarrow$ & $-56.6$\% &3.80& -53.2\% & 8.45 & -55.5\% & 44.1 \\
ts$\leftrightarrow$ & $-5.4\%$ & 3.85  & -7.22\% & 8.96 &-10.6\% & 46.6\\
\hline
\end{tabular}
\end{center}
\caption{\label{table:stats}The average percentual change in tick size ($\Delta$tick size) and the average market capitalization in billions of euros, of the nine defined groups.}
\end{table}

In this article a couple of quantities is considered in order to asses the influence of the MiFID II tick size regime. Every quantity is computed as a per day average for every stock, after which it is averaged over the year to find the average per stock per year.
Then the percentual change in the value of the quantity in 2018 (after the introduction of the MiFID II tick size regime) relative to the value of the quantity in 2017 (before the introduction of MiFID II) is computed for every stock and this is also averaged over the different groups. This is the level on which statistical significance is assessed: a standard t-test and a Wilcoxon signed rank test\footnote{The Wilcoxon signed rank test is added because it does not assume normality of the samples.} are conducted to test the null-hypothesis that the mean percentual change of the considered group (ts$\uparrow$ or ts$\downarrow$) equals the mean percentual change of the control group (ts$\leftrightarrow$) versus the alternative hypothesis that the mean percentual changes of the both groups differ. 
Furthermore, to complement the pairwise comparison of the pre- and post-MiFID II periods, we run a panel regression including stock fixed effects. That is, for a certain dependent variable of interest DepVar, we run the following regression,
\begin{align} 
\label{eq:panel_regression}
\text{DepVar}_{i,t} &= \beta_0+ \beta_1 \mathbf{1}_{+}(i)\mathbf{1}_{\text{M}}(t) + \beta_2 \mathbf{1}_{-}(i)\mathbf{1}_{\text{M}}(t) + \beta_3 \mathbf{1}_{+}(i) + \beta_4 \mathbf{1}_-(i)+ \beta_5 \mathbf{1}_{\text{M}}(t) \nonumber \\
&\quad+ \text{Control Vars}_{i,t} + \text{SF}_i+\epsilon_{i,t},
\end{align}
where the subscript $i,t$ indicates stock $i$ on day $t$ (running over all trading days in 2017 and 2018). Here, $\mathbf{1}_{\text{M}}(t)$ equals 1 when $t$ is after MiFID II (\ie\ $t$ is in 2018) and 0 otherwise and $\mathbf{1}_{+}(i)$ ($\mathbf{1}_{-}(i)$) equals 1 when stock $i$ is in the group of which the tick size was increased (decreased), and 0 otherwise. Furthermore, Control Vars denotes a set of control variables. Following \cite{Harris94}, we include market capitalization and daily transacted volume (both in log-scale),  stock price (measured on the close) and volatility (measured as the daily sum of squared one minute mid price returns) as control variables. Finally, SF$_i$ denotes the stock fixed effect of stock $i$.
We are primarily interested in (the sign of) the coefficients $\beta_1$ and $\beta_2$, because these capture the effects of an increase or decrease in tick size post-MiFID II. Furthermore, $\beta_5$ is interesting as it captures effects that are observed since the introduction of MiFID II that can not be attributed to the new tick size regime, neither to one of the control variables, nor to stock fixed effects. These are (indirect) effects of other MiFID II regulations, which we discuss in section \ref{sec:closing auctions}. 

Finally, it is known that price formation in closing auctions is highly influenced by calendar effects. Firstly, transacted volume in closing auctions is significantly higher on the third friday of the month, when options expire and indices are reweighted. Also, transacted volume in closing auctions is higher on the last trading day of the month, when MSCI indices are reweighted and window dressing effects can be expected. See table \ref{table:calendar_effects} for an illustration of these effects. For those reasons, we exclude third fridays and last trading days of the month from our analysis, to only take `normal' market conditions into account.
\begin{table}[h]
\small
\begin{center}
\begin{tabular}{ p{2.5cm} p{1.cm} p{1.cm} p{1.cm}}
 \hline
 \multicolumn{4}{c}{Transacted volume in closing auctions} \\
Day &\multicolumn{1}{c}{Small cap stocks} & \multicolumn{1}{c}{Mid cap stocks} &\multicolumn{1}{c}{Large cap stocks} \\
\hline
Third friday & 6.64  & 12.41 & 50.85 \\
End of month & 6.59 & 13.50 & 46.74 \\
`Normal' day & 3.21 & 6.61 & 24.93 \\
\hline
\end{tabular}
\end{center}
\caption{\label{table:calendar_effects} Average transacted volumes in closing auctions (measured in millions of euros), for special days (third friday and last trading day of the month) versus normal days.}
\end{table}

\section{Effects of tick size on liquidity and transacted volume}
\label{sec:liqudity}
In this section we discuss the effects of the new tick size regime on liquidity and transacted volume. In particular, we test whether Intraday Hypotheses \ref{hyp:intraday_liq} and \ref{hyp:intraday_tv} and Closing Auction Hypotheses \ref{hyp:auction_liq} and \ref{hyp:auction_tv} hold true.
\subsection{Liquidity measures}
First, we will discuss the  different liquidity measures we use, starting with intraday liquidity measures.
The quoted \emph{bid-ask spread} at time $t$ is defined as 
$$\text{Bid-ask spread} = \frac{BA_t- BB_t}{MP_t},$$
where $BA_t$ and $BB_t$ are the best ask and best bid price in the book at time $t$ and $MP_t=\frac12(BA_t+BB_t)$.
The \emph{effective spread} for a trade $T$ at time $t$ is given by
$$\text{Effective spread}=\frac{2I_T(P_T-MP_t)}{MP_t},$$
where $P_T$ is the price of trade $T$ and $I_T$ is an indicator equal to $\pm1$ if the trade $T$ is buyer/seller initiated.\footnote{Following for example \cite{Bessembinder03} and \cite{Chakravartyetal04}, we include the effective spread because the quoted bid-ask spread does not account for transactions that occur inside the spread.} Furthermore by \emph{bid-ask vol1} we denote the total volume (in euros) on the best bid and ask, and by \emph{bid-ask vol3} (\emph{bid-ask vol5}) we denote the total volume in euros on the first 3 levels (5 levels) on both sides of the limit order book. To measure spread and depth jointly, we consider the (theoretical) price impact of a market order. The \emph{price impact x} of an order of size $x$ (in euros) on time $t$ is defined as 
\begin{align} \label{eq:price_impact}
\text{Price impact } x = \frac{(A^x_t - MP_t) + (MP_t-B^x_t)}{2MP_t}, 
\end{align}
where $A^x_t$ is the lowest ask price in the order book such that the cumulative ask volume (in euros) for a price lower than or equal to $A^x_t$ is greater than or equal to $x$ and $B^x_t$ is the highest bid price in the order book such that the cumulative bid volume (in euros) for a price greater than or equal to $B^x_t$ is greater than or equal to $x$.  This variable measures the price deviation that a market order of size $x$ would cause (in fact, the average of the price impact of a buy and a sell market order). Note that for $x=0$ it is equal to half the bid-ask spread, so for low values of $x$ this measure is mainly driven by the bid-ask spread, while for higher values of $x$ the volumes in the order book are more important.\footnote{This measure is essentially equal to a half times the cost of round trip trade (CRT) as defined in \cite{Domowitzetal05} and \cite{Irvineetal00}, however we choose to measure a market order in euros instead of in number of shares, because share prices vary a lot within our sample.}
All quantities are measured every minute and averaged over the day, except the effective spread, which is measured for every trade and turned into a daily trade size weighted average.

To investigate liquidity in the closing auction, we need different measures. By the very nature of the call auction mechanism, there is no bid-ask spread (as bid and ask prices overlap during the auction). However, the \emph{post auction bid-ask spread}, defined as the bid-ask spread in the book just after the clearing of the auction, serves as an indicator of how scattered liquidity is.
By \emph{bid-ask vol1} we denote the volume (in euros) on the best bid and ask just after clearing, and by \emph{bid-ask vol3} (\emph{bid-ask vol5}) we denote the volume in euros on the first 3 levels (5 levels) on both sides of the limit order book just after clearing.  To measure the tradeoff between post auction spread and depth, one would ideally want to compute the price impact of submitting an extra market order to the auction, similar to the price impact measure for intraday trading. However, to determine the real theoretical price impact of submitting an extra market order to the auction, one would need the whole overlapping order book of the auction, which we do not have. Instead we employ Amihud's illiquidity ratio \citep{Amihud02} to measure price impact, defined as the absolute return divided by the transacted volume. In order to use this measure for the closing auction, we define the \emph{closing auction return} $R^{CA}_t$ on day $t$ by
\begin{align} \label{eq:def_returns} R^{CA}_t = \log X^{CA}_t - \log \bar{X}_t,
\end{align}
where $X_t^{CA}$ denotes the auction's closing price on day $t$ and $\bar{X}_t$ is the volume weighted average price (VWAP) over the last five minutes of continuous trading. Furthermore, we denote by $TV^{CA}_t$ the transacted volume (in millions of euros) in the closing auction for day $t$. Then the illiquidity ratio on day $t$ equals,
\begin{align} \label{eq:amihud}
\text{Illiquidity} = \frac{|R^{CA}_t|}{TV^{CA}_t}.
\end{align}
Note that this quantity measures the price change per unit of transacted volume, and thus is a measure of illiquidity. We use this measure to quantify price impact in closing auctions, which is justified by results of \cite{Goyenkoetal09}, who find that the illiquidity ratio is a proper measure of price impact.
\subsection{Results for liquidity}
Table \ref{table:liquidity} shows the intraday liquidity measures
for every group pre- and post-MiFID II and the average increase of the measure per group. The results suggest that an increase (decrease) in tick size leads to an increase (decrease) in bid-ask spread (both quoted and effective) and an increase in the volumes in the limit order book (confirmed up to 5 levels deep). 
This observation is confirmed by the regression results, which are shown in table \ref{table:liquidity_regression}. The coefficient $\beta_1$ for an increase in tick size ($\beta_2$ for a decrease in tick size) is found to be significantly positive (negative) for bid-ask spread, effective spread and order book volumes for all market cap groups. To check if an increase in depth can offset the increase in spread, we look at the theoretical price impact. The coefficients $\beta_1$ (corresponding to an increase in tick size) in table \ref{table:liquidity_regression} for price impacts are significantly positive for market orders up to 50,000 euros for large caps and mid caps and up to 10,000 euros for small caps, while the coefficients $\beta_2$ for price impacts are all significantly negative. This indicates that the increase in depth is not able to offset the increase in bid-ask spread completely (only for larger orders). We conclude that an increase (decrease) in tick size has a negative (positive) effect on intraday liquidity, except for large orders. This implies that Intraday Hypothesis \ref{hyp:intraday_liq} is confirmed for the most part. However, we do not find a decrease in liquidity for large orders after a tick size reduction, hence the decrease in spread compensates for the decrease in depth in this case.
\begin{center}
[Insert tables \ref{table:liquidity} and \ref{table:liquidity_regression} here.]
\end{center}

Results are different when liquidity in the closing auction is considered. In table \ref{table:liquidity_auc} descriptive statistics and in table \ref{table:liquidity_regression_auc} regression results for the closing auction liquidity measures are shown. For large cap stocks, an increase in tick size is associated with only a small increase in post auction bid-ask spread, as is shown by the fact that $\beta_1=0.451$, much lower than the effect observed in intraday trading ($\beta_1=2.251$, table \ref{table:liquidity_regression}).  On the other hand, the coefficient $\beta_2$ for large caps whose tick size decreased equals $-2.133$, which is more comparable to the intraday continuous trading effect for large caps ($\beta_2=-2.589$, table \ref{table:liquidity_regression}). Similar (but even stronger) effects are observed for mid and small caps, where the post auction spread does not increase after an increase in tick size. The differences with results for continuous trading are explained by the observation that post auction spreads are in general higher than intraday bid-ask spreads (compare the values for bid-ask spread in table \ref{table:liquidity} and post auction spread in table \ref{table:liquidity_auc}), causing that the tick size is rarely a binding constraint for the post auction spread.\footnote{To illustrate this, in 2017 only 10.96\% of the stocks in our sample had an average post auction spread (over the year 2017) below 1.25 tick sizes, while 25.05\% of the stocks had an average intraday bid-ask spread below 1.25 tick sizes (in both cases, these stocks only belonged to the groups ts$\leftrightarrow$ or ts$\downarrow$). In 2018, these percentages changed to $2.99\%$ for post auction spreads and $4.88\%$ for intraday spreads indicating the new tick size regime lead to more suitable tick sizes that are less binding the spread.} Although results for the spread differ between intraday and closing auction trading, results for depth are very similar. Also in case of auction trading, volumes in the order book increase (decrease) following an increase (decrease) in tick size, as can be seen from the significantly positive values for $\beta_1$ (and significantly negative $\beta_2$) in table \ref{table:liquidity_regression_auc} for all market cap groups. 
 The fact that depth strongly increases with an increase in tick size, while the post auction spread is largely unaffected, suggests that price impact decreases (and thus liquidity improves) when the tick size increases. This is confirmed by the results for the illiquidity ratio: for large, mid and small cap stocks the $\beta_1$ is significantly negative, showing that closing auction liquidity improves when the tick size increases. For large and mid caps, $\beta_2$ is not significantly different from zero, indicating that the decrease in depth following a tick size decrease did not harm liquidity. However, for small caps the $\beta_2$ for illiquidity is significant and positive, showing a decrease in tick size negatively affected closing auction liquidity. We conclude that Closing Auction Hypothesis \ref{hyp:auction_liq} holds true for the most part: indeed depth is positively (negatively) affected by an increase (decrease) in tick size, but post auction spreads do not increase with an increase in tick size (only slightly for large caps). This leads to an improvement in closing auction liquidity after an increase in tick size, but negative effects of a decrease in tick size are only observed for small cap stocks.
\begin{center}
[Insert tables \ref{table:liquidity_auc} and \ref{table:liquidity_regression_auc} here.]
\end{center}
\subsection{Results for transacted volume}
The results in table \ref{table:liquidity_regression} on transacted volume show that an increase (decrease) of the minimum tick size has a positive (negative) effect on the intraday transacted volume. For all market cap groups the coefficient $\beta_2$ for a decrease in tick size is significantly negative and for small and large caps the coefficient $\beta_1$ for an increase in tick size is significantly positive. 
So we find evidence that intraday transacted volume increases with an increase in tick size, and vice versa, causing us to reject Intraday Hypothesis \ref{hyp:intraday_tv}. A possible explanation of this observation is that the higher transacted volumes are caused by high frequency traders (HFT). Several studies (see \eg\ \cite{Hagstromernorden13,Oharaetal18,Yaoye18}) show that a higher tick size is more attractive for HFT-strategies, as it implies a higher bid-ask spread and increases the importancy of time priority. This possibly caused a shift of HFT activity towards stocks with a high tick size in the new regime.  Another explanation is the simple mechanical argument by \cite{ChiarellaIori02}, who argue that a higher tick size causes more orders to collapse into the same price level, leading to higher transacted volumes. 

As already discussed in section \ref{sec:hypos}, this latter argument is more appealing in case of the call auction mechanism, where less interaction is possible and this type of mechanical effects should be stronger. Indeed, we find that an increase (decrease) in tick size is associated with an increase (decrease) in transacted volume in the closing auction, as is shown by the significantly positive values for $\beta_1$ and significantly negative values for $\beta_2$ in table \ref{table:liquidity_regression_auc}, for transacted volume. Also, the effect is stronger in the closing auction than for intraday continuous trading (as is seen from the coefficients that are higher in absolute value for the closing auction), supporting the abovementioned mechanical argument. We conclude that Closing Auction Hypothesis \ref{hyp:auction_tv} holds true. Finally, it is noteworthy that the post-MiFID II coefficient $\beta_5$ in table \ref{table:liquidity_regression_auc} is highly positive and significant for transacted volume in closing auctions, showing that auction volumes increased heavily since MiFID II, independent of the new tick size regime, a fact that is investigated in more depth in section \ref{sec:closing auctions}.

\section{Effects of tick size on volatility and market stability} \label{sec:stability}
In this section we examine the effects of the new tick size regime on volatility and market stability, by testing whether Intraday Hypotheses \ref{hyp:intraday_vola} and \ref{hyp:intraday_stability} and Closing Auction Hypotheses \ref{hyp:auction_vola} and \ref{hyp:auction_stability} hold true.
\subsection{Results for volatility} \label{sec:vola}
In panel A of table \ref{table:volatility and market stability} intraday volatility and market stability measures are shown. Here, volatility is measured as the daily sum of squared one minute mid price returns. The values suggest a positive relationship between volatility and tick size. This is partially confirmed by the regression results that are shown in panel A of table \ref{table:stability_regression}: for large cap stocks the coefficient $\beta_1$ (associated with an increased tick size) is significantly positive and $\beta_2$ (for a decrease in tick size) is significantly negative, indicating that tick size and volatility are indeed positively related. However, for mid cap stocks this relationship is only significant for a decrease in tick size and for small cap stocks only for an increase in tick size. Moreover, although the coefficients are statistically significant, they are not economically: for example, an increase in tick size is for large cap stocks only associated with a 4.5 basis points increase in daily volatility. So we conclude that  Intraday Hypothesis \ref{hyp:intraday_vola} holds true for large cap stocks and only partially for mid and small cap stocks, but that observed effects are minimal.

Similar effects are observed when we consider the closing auction. To measure price deviations in the closing auction, we look at the absolute value of the closing auction return (as defined in equation (\ref{eq:def_returns})).
The absolute auction return is weakly affected by the new tick size regime: the results in panel B of table \ref{table:stability_regression} show that higher (lower) tick sizes lead to higher (lower) absolute auction returns, indicated by positive values for $\beta_1$ and negative values for $\beta_2$ in table \ref{table:stability_regression}, for all market cap groups. However, $\beta_1$ is only significant for large cap stocks and $\beta_2$ only for large and mid cap stocks. We conclude that Closing Auction Hypothesis \ref{hyp:auction_vola} only holds true for large caps and that observed effects are rather weak. Noteworthy, the values for $\beta_5$ for absolute auction returns in panel B of tabel \ref{table:stability_regression} are all highly positive and significant, indicating closing price deviations have increased substantially since MiFID II, a fact that is investigated more thoroughly in section \ref{sec:closing auctions}.
 \begin{center}
 [Insert tables \ref{table:volatility and market stability} and \ref{table:stability_regression} here.]
\end{center}
\subsection{Results for market stability}
To create a more stable market microstructure (which is one of the goals of MiFID II), a lower number of bid-ask price changes and a lower number of (small) trades, combined with a higher trade size are desired. The results in panel A of table \ref{table:volatility and market stability} indicate that for small, mid and large cap stocks an increase (decrease) in tick size has a negative (positive) effect on the number of best bid-ask price changes.
This is confirmed by the regression results in panel A of table \ref{table:stability_regression}, where it is shown that the $\beta_1$ for the number of bid-ask price updates is significantly negative for all market cap groups and the $\beta_2$ is significantly positive. A similar effect we observe for the closing auction, where 
the number of indication price updates significantly decreases (increases) with an increase (decrease) in tick size, as is seen from the negative and significant coefficients $\beta_1$ (and the positive and significant coefficients $\beta_2$), for all market cap groups in panel B of table \ref{table:stability_regression}.
 
 Another important metric to measure the activity in the market, is the number of intraday trades. We document a negative relationship between tick size and the number of trades, as the coefficient $\beta_1$ for the number of trades in table \ref{table:stability_regression} is negative and significant for all market cap groups, while $\beta_2$ is positive and significant. This does not imply a negative relationship between tick size and transacted volume (\cf\ section \ref{sec:liqudity}). It just means that the average size of trades increases with an increase of the minimum tick size, as becomes clear from the average trade size values in panel A of table \ref{table:volatility and market stability} and the regression coefficients in panel A of table \ref{table:stability_regression} that are significantly positive (negative) for an increase (decrease) in tick size. In light of the discussed results, we conclude that an increase in tick size positively affects the stability of the market, while a decrease in tick size has an adverse effect on market stability, for both intraday continuous trading and the closing auction. It is thus concluded that Intraday Hypothesis \ref{hyp:intraday_stability} and Closing Auction Hypothesis \ref{hyp:auction_stability} both hold true.

\section{MiFID II effects beyond the new tick size regime} \label{sec:closing auctions}
In this section post-MiFID II effects that can not be attributed to the new tick size regime are discussed. To detect those effects, we consider the values for $\beta_5$ in tables \ref{table:liquidity_regression}, \ref{table:liquidity_regression_auc} and \ref{table:stability_regression}, because the coefficient $\beta_5$ is associated with post-MiFID II effects that are not explained by the new tick size regime, stock fixed effects or one of the control variables. Although some of the values for $\beta_5$ for intraday quantities in table \ref{table:liquidity_regression} and panel A of table \ref{table:stability_regression} are statistically significant, none of them is relatively large (compared to values of $\beta_1$ and $\beta_2$, or to average values in tables \ref{table:liquidity} and \ref{table:volatility and market stability}), indicating that MiFID II effects not caused by the tick size regime are minimal for intraday continuous trading. On the other hand, both statistically and economically significant effects are observed in the closing auction, which is indicated by relatively high values for $\beta_5$ in table \ref{table:liquidity_regression_auc} and panel B of table \ref{table:stability_regression}. Most remarkable, transacted volumes and liquidity in the closing auction increased heavily, as well as closing price deviations. This suggests that the closing auction is more heavily affected by the other MiFID II regulations than intraday continuous trading.  In this section, we shed some light on these effects.
\subsection{Transacted volume in the closing auction}
\begin{figure}[h]
\begin{center}
\parbox{13.5cm}{
\includegraphics[scale=0.55]{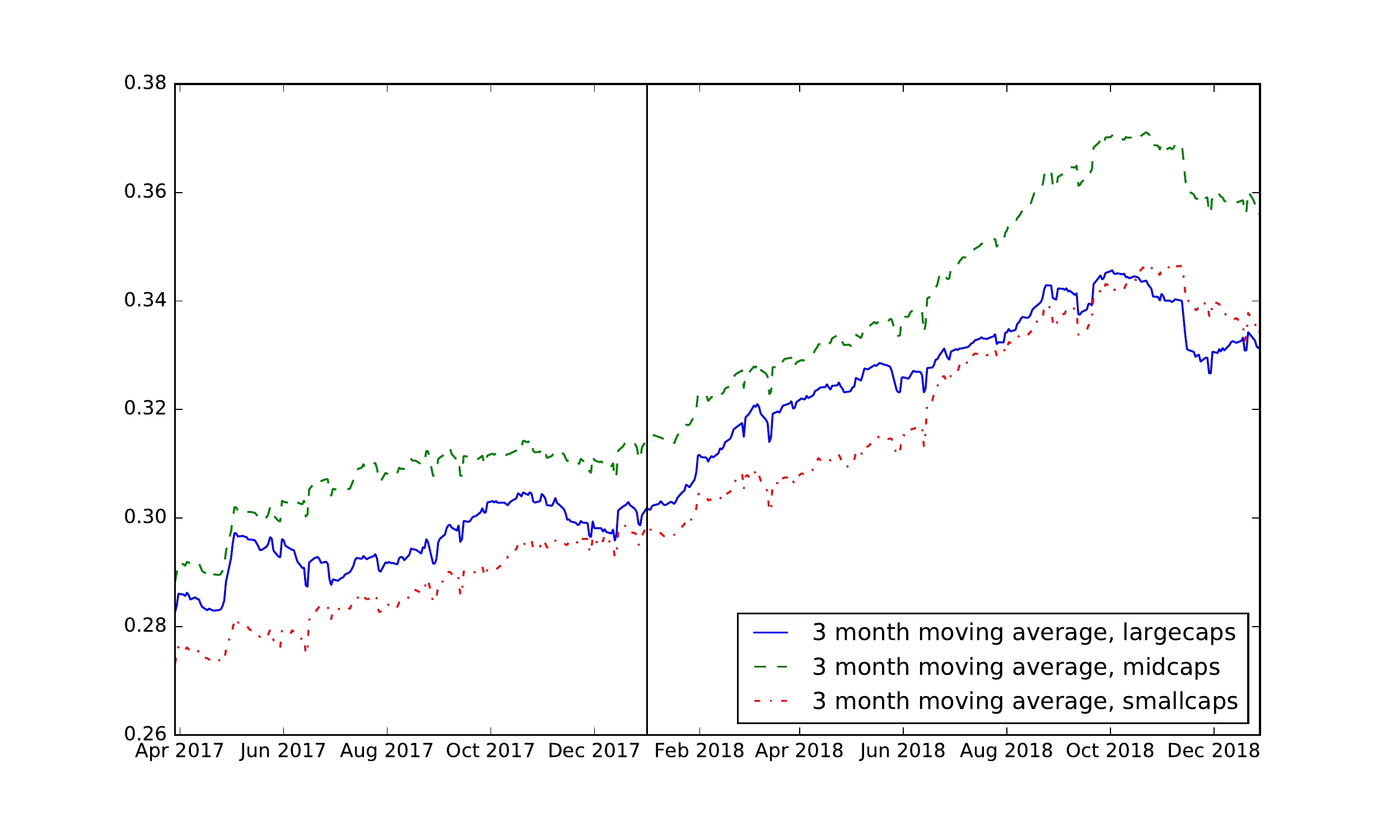}\caption{How the closing auction volume/total transacted volume-ratio increased in 2018 with respect to 2017. The vertical black line denotes the introduction of MiFID II on January 3rd 2018.}
\label{fig:closing_auction_vol}
}
\end{center}
\end{figure}

The last column of table \ref{table:liquidity_regression_auc} shows that the $\beta_5$ coefficient is both statistically and economically significant for transacted volume in the closing auction. For example, for large cap stocks $\beta_5$ equals $0.229$, indicating an additional increase in auction volume of $22.9\%$ since MiFID II that can not be explained by tick size effects, stock fixed effects or control variables such as volatility. For mid and small cap stocks even higher values for $\beta_5$ are reported. While auction volume has increased strongly since MiFID II, intraday volume has only increased slightly, as can be seen from the lower values for $\beta_5$ of transacted volume in table \ref{table:liquidity_regression}. 
These two effects together lead to a significant increase in the fraction of total volume that is transacted in the closing auction. In figure \ref{fig:closing_auction_vol} it is illustrated that this increase in importance of the closing auction already started before 2018, but was amplified after the introduction of MiFID II (the vertical black line in the figure). 

There are a few possible explanations for this observation.  
In particular there are two parts of MiFID II that seem reasonable causes of this shift towards the closing auction: the restrictions on dark pools (DVC rule) and the \emph{best execution} rule. As trading in the closing auction is far more similar to dark pool trading than continuous trading, a reasonable explanation of the observed effect would be that volume moved from dark pools to closing auctions. However, our analysis in appendix \ref{appendix:darkpools} shows that there is no evidence that this is the case, in line with \cite{Johannetal19}. Under the best execution rule, intermediaries are obliged to execute a client's order against the best possible price. Moreover, their obligations to report the details of the execution have increased strongly in MiFID II regulations. Interestingly, trading in the closing auction makes fulfillment of this reporting obligation a lot easier, since the closing auction offers a single price against which all transactions are executed. Also, during the closing auction there are no transactions, so it is unclear what the best price is.  As it is impossible for market participants to determine the best price, they might refrain from trading (on other venues) during the closing auction and instead wait to trade on the close.

\begin{figure}[h]
\begin{center}
\parbox{13.5cm}{
\includegraphics[scale=0.45]{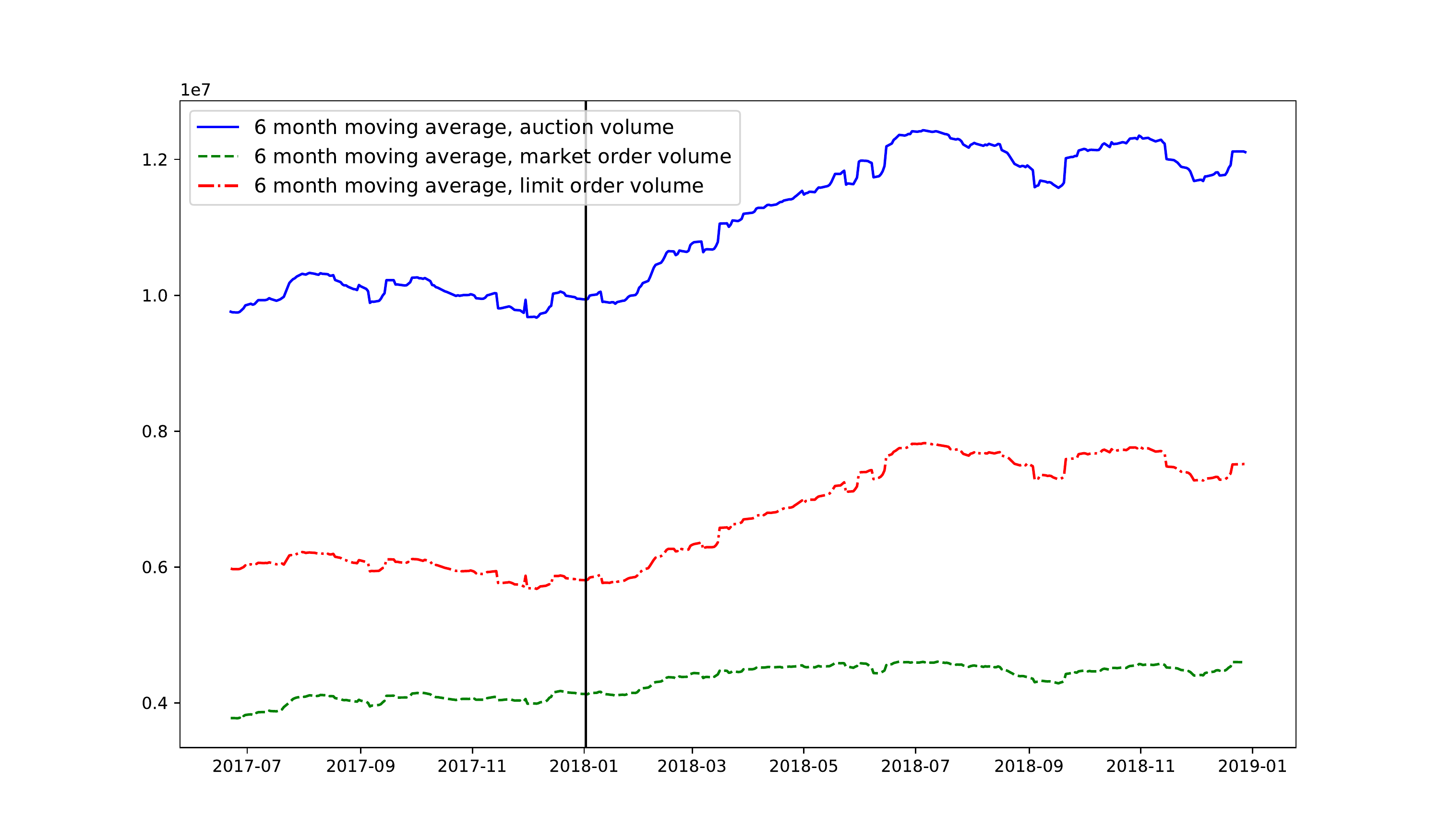}\caption{How the transacted volume (in euros) and the executed market- and limit-order volumes (in euros) increased since the introduction of MiFID II (the solid black line denotes the introduction of MiFID II). The figure shows moving averages computed over all STOXX600 constituents traded on Euronext. Clearly, the increase in auction volume is mainly driven by an increase in executed limit order volume}
\label{fig:market_vs_limit}
}
\end{center}
\end{figure}

Independent of MiFID II, the increase of volumes in the closing auction is already going on for some time and is related to the increase of passive investing in the past years and the growing use of exchange traded funds (ETFs) in particular. The increase in passive investing is associated with a higher amount of market-on-close orders, consistent with an aim for minimal tracking errors\footnote{A market-on-close order is a market order that is executed in the closing auction, it thus automatically serves for the need of minimal tracking error, as the order is for sure executed against the benchmarking closing price.} (see \cite{Bogousslavskymuravyev19} and \cite{Wu19}). So, if the increase in closing auction volume is mainly caused by the increase in passive investing, it should be driven by an increase in market order volume. However, figure \ref{fig:market_vs_limit} shows this is not the case: the increase in volume is mainly driven by an increase in executed limit order volume. 
Notice that market-on-close volume does not contribute to price formation much, as market orders on both sides of the market are executed against each other without affecting the price formation process. This is in line with \cite{Bogousslavskymuravyev19} and \cite{Wu19}, who show that increased passive investing and the associated market-on-close volume lead to inefficient closing prices. Also, in the setting of continuous trading, \cite{Bendavidetal18} show that high ETF ownership is associated with less efficient prices. In the next subsections we examine efficiency of price formation in the closing auction, pre- and post-MiFID II, in order to find out if the observed extra volume contributes to more efficient price formation.

\subsection{Efficiency of the price formation process}
We already highlighted in section \ref{sec:vola} that absolute auction returns have increased since MiFID II, as pointed out by the significant and positive values of the post-MiFID II coefficient $\beta_5$ for absolute auction returns in panel B of table \ref{table:stability_regression}. For all market cap groups, the coefficient $\beta_5$ is around 3 basis points (bps), indicating a post-MiFID II increase in closing price deviations by 3 bps, that can not be explained by the tick size regime, stock fixed effects or intraday volatility.
From this, one could at first sight naively conclude that price formation in closing auctions has become less efficient. Indeed, higher absolute returns could indicate more noise in the closing prices and hence a less efficient price formation process. However, if the higher returns were generally in the right direction (that is, towards the `efficient price'), it would actually mean that price formation in closing auctions became more efficient. In order to decide which of the two is the case, we look at the mean reversion the day after an auction with a high absolute return  (higher than 10 bps).\footnote{Reported results do not radically change if this boundary is changed to \eg\ 5 bps or 30 bps, but we want to exclude auctions with very low returns, as reversion is not well-defined when a return close to 0 is made in the auction.} If the next day a correction takes place, this is a sign that the closing price was not efficient. To make this rigorous, we define the   \emph{overnight return} $R_{t,t+1}$ from day $t$ to day $t+1$ by
\begin{align} R_{t,t+1} = \log \bar{X}_{t+1} - \log X_t^{CA},
 \end{align}
 where $X_t^{CA}$ denotes the closing price on day $t$ and $\bar{X}_{t+1}$ is the volume weighted average price over the last five minutes of continuous trading on day $t+1$.
Note that this is the return made from the closing price on day $t$ to the last five minutes of continuous trading before the closing auction on day $t+1$.\footnote{We differ in this definition from other studies, where the overnight return is measured from close on day $t$ to open on day $t+1$ (see \eg\ \cite{Louetal18} or \cite{Hendershottetal20}), because the opening auction is not comparable to the closing auction at all in terms of transacted volume. In the closing auction on average around 30\% of the daily volume is transacted, while the opening auction accounts for less than 1\% on average. Results are similar but less distinctive, when we instead measure the overnight return until open.}
We estimate the following simple linear regression model,
\begin{align} \label{eq:regression} R_{t,t+1} = c + b R^{CA}_t + \epsilon_t,
 \end{align}
 for all days $t$ with an absolute auction return $|R^{CA}_t|$ above 10 bps, where the auction return $R^{CA}_t$ is defined as in equation (\ref{eq:def_returns}).
The quantity of interest is the coefficient $b$, which measures if overnight mean reversion takes place: if $b$ is significantly below zero, it means that the returns made in the closing auction are (partially) reverted the next trading day. If $b$ is not significantly different from zero, it means that there is no linear correlation between the closing auction return and the following overnight return. In the first column of table \ref{table:corrections_ts} the estimated values of $b$ are displayed, over 2017 and 2018, for small, mid and large cap stocks. The results make clear that over 2017 closing auction returns and following overnight returns were significantly negatively correlated, meaning that large absolute returns in the closing auction were partially corrected the next day. However, this effect completely vanished in 2018, as it is seen that the mean reversion coefficient $b$ is not significantly different from zero for small, mid and large caps. It is thus concluded that closing prices have become efficient in 2018, while they where significantly inefficient in 2017. The other columns in table \ref{table:corrections_ts} show the results when we double sort the stocks for tick size groups, making clear that the observed effects are not related to the new tick size regime.
\begin{table}[h]
\footnotesize
%\begin{center}
\begin{tabular}{ p{3. cm} p{1. cm} p{1. cm}p{0.001cm} p{1. cm}p{1. cm} p{0.001cm} p{1. cm} p{1. cm} p{0.001cm} p{1.cm} p{1. cm}  }
 \multicolumn{12}{c}{Mean reversion coefficients}\\
 \hline \\
  &\multicolumn{2}{c}{All} &&\multicolumn{2}{c}{ts$\leftrightarrow$(control group)}& & \multicolumn{2}{c}{ ts$\uparrow$ }& &\multicolumn{2}{c}{ts$\downarrow$} \\
 \cline{2-3} \cline{5-6} \cline{8-9} \cline{11-12}\\
&2017 & 2018  && 2017 & 2018  &&2017 & 2018 &&2017 & 2018 \\
%\cline{2-12} 
  \cline{2-3} \cline{5-6} \cline{8-9} \cline{11-12} \\
Large caps & $-0.297^{*}$ & 0.0265 && $-0.141^{*}$ & 0.0092 &&  $-0.402^*$ & 0.0129 && $-0.329^*$ &  0.125 \\
Mid caps & $-0.286^{*}$ & $-0.00996$ && $-0.175^*$ & 0.0120 && $-0.449^*$ & $0.000212$ && $-0.2302^*$ & 0.119 \\
Small caps & $-0.203^{*}$ & 0.0658 && $-0.1492^*$ & 0.1416&&  $-0.367^*$ & $-0.171^*$ && $-0.1371^*$ & $0.2498^*$ \\
\hline
\end{tabular}
%\end{center}
\caption{\label{table:corrections_ts} 
How the mean reversion coefficient $b$ of the model in equation (\ref{eq:regression}), changed in 2018 compared to 2017, for auctions with absolute returns higher than 10 bps.\\
\footnotesize{* The mean reversion $b$ is significantly different from 0 at the $0.01$-level.}}
\end{table}

\subsection{Volume and efficiency}
We reported that along with the increased volumes in 2018, the price deviations in the closing auctions have increased, while mean reversion on the next day vanished. Usually higher transacted volumes are associated with more efficient price formation. Here, we investigate to which extent the two effects are related. To do so, we estimate for both years the following regression model,
\begin{align} \label{eq:regression_vol} R_{t,t+1} = c_1 + c_2\mathbf{1}_{Q1} + c_3\mathbf{1}_{Q4} + b_1 R_t^{CA} + b_2 R_t^{CA} \mathbf{1}_{Q1} + b_3 \mathbf{1}_{Q4}R_t^{CA} + \epsilon_t, 
\end{align}
where $\mathbf{1}_{Qi},i=1,4$ denotes if a day is in the lowest ($Q1$) or highest ($Q4$) quartile of closing auction volumes for the particular stock and year. The coefficients of interest are again the parameters $b_i, i=1,2,3$. Now $b_1$ denotes the mean reversion coefficient for days with median volumes (\ie\ all days that fall in the second and third quartile of closing auction volumes for the particular stock). Furthermore, $b_2$ denotes the \emph{additional} mean reversion that is observed for days with low volumes, meaning the corresponding mean reversion coefficient equals $b_1+b_2$. Similarly, $b_1+b_3$ is the mean reversion coefficient for days with high volumes. In table \ref{table:role_volume_mr} the results of this regression are summarized. The reported values are the mean reversion coefficients for the three groups, \ie\ $b_1+b_2$ for Q1, $b_1$ for Q2,3 and $b_1+b_3$ for Q4.  It is seen from table \ref{table:role_volume_mr} that there exists some relationship between volume and efficiency: in 2017 the high (low) volume days are associated with less (more) overnight mean reversion, although the differences are not all statistically significant. Importantly, the mean reversion coefficient is (except for small cap stocks) still below 0 for the high volume days in 2017. This in contrast to the results for 2018 in table \ref{table:role_volume_mr}: there is no significant mean reversion after a high auction return for any of the groups (Q1, Q2,3 or Q4). This implies that the vanishing of the mean reversion in 2018 can not be fully explained by the increase in transacted volumes. 

To summarize, there exists a positive relationship between auction volume and closing price efficiency, but this is not strong enough to account for the remarkable differences between 2017 and 2018. This indicates that the extra volume that has come to the closing auctions since MiFID II is of a different nature, coming from market participants that primarily use limit orders and contribute to more efficient price formation. The results are consistent with the conjecture that MiFID II regulations (possibly by means of the best execution rule) attract professionals to the closing auction on Europe's primary exchanges. On the other hand, results are at odds with an explanation that links the increased volume to the increased passive investing.
 \begin{table}[h]
\footnotesize
\begin{center}
\begin{tabular}{  p{0.7cm} p{1.25cm}p{1.18cm}p{1.18cm} p{1.25cm}p{1.18cm}p{1.25cm} p{1.5cm}p{1.18cm}p{1.25cm} }
 \multicolumn{10}{c}{Mean reversion coefficient for high and low volume days} \\
 \hline
 &\multicolumn{3}{c}{Small cap stocks} & \multicolumn{3}{c}{Mid cap stocks} &\multicolumn{3}{c}{Large cap stocks} \\
 & Q4 & Q2,3 &Q1  & Q4 & Q2,3 & Q1 & Q4 & Q2,3 & Q1 \\
2017 & 0.0629$^{+}$ &-0.323$^{*}$ & -0.422$^*$ & -0.206$^*$ & -0.299$^{*}$ & -0.452$^*$ &-0.095$^{*+}$ & -0.367$^*$ & -0.497$^*$ \\
2018
&0.109 &0.0488  & 0.0286& -0.0845& 0.0647& 0.172 & -0.0513 & 0.149 & -0.0495 \\
\hline
\end{tabular}\\
\end{center}
\caption{\label{table:role_volume_mr}
How the mean reversion coefficient  of the model in equation (\ref{eq:regression_vol}), changed in 2018 compared to 2017. The mean reversion coefficient $b_1$ is the value reported under Q2,3, the value reported under Q4 is $b_1+b_3$ and the value under Q1 is $b_1+b_2$, as $b_2$ and $b_3$ give the difference in mean reversion coefficient between Q2,3 and Q1, Q4 respectively.   \\ \footnotesize{* The mean reversion coefficient ($b_1$ for Q2,3, $b_1+b_2$ for Q1, $b_1+b_3$ for Q4) is significantly different from 0 on the 0.05-level. \\ 
+ The mean reversion coefficient $b_1+b_i,~i=2,3$ is significantly different from the mean reversion coefficient $b_1$ belonging to the Q2,3 group in the same year and market cap, on the 0.05-level.} }
\end{table}
\subsection{MiFID II or other factors: checking the S\&P500}
As a robustness check for the above arguments linking MiFID II and the closing auction effects, we verify that observed effects do not appear outside of the MiFID II purview. For this purpose, we obtain closing auction data of 250 randomly selected stocks belonging to the S\&P500, for the years 2017 and 2018. In table \ref{table:auctions_sp500} the average transacted volume in closing auctions is shown, as well as the average absolute return made in the closing auctions, where the closing auction return is again defined as in equation (\ref{eq:def_returns}). Clearly, we see a significant increase in the  closing auction's transacted volume and along with this increase in volume, a similarly sized increase in the average absolute returns. This increase in volumes and deviations is linked to the increase in passive investing \citep{Bogousslavskymuravyev19,Wu19}.
\begin{table}[h]
\footnotesize
\begin{center}
\begin{tabular}{ p{5 cm} p{1.5 cm} p{1.5 cm} p{1.5 cm} }
 \multicolumn{3}{c}{Closing auction statistics in the US}\\
 \hline \\
Variable &2017 & 2018 &increase \\
\hline\\
Transacted volume & 23.92 & 37.18  & $55.43\%^{*}$ \\
Absolute auc return & 3.87 & 6.04 & $56.07\%^{*}$ \\
Mean reversion coeff. $b$ (10 bps)  & $-1.25^{**}$ & $-0.693^{**}$ &$-$ \\
Mean reversion coeff. $b$ (30 bps)  & $-1.28^{**}$ & $-1.54^{**}$ &$-$ \\
\hline
\end{tabular}
\end{center}
\caption{\label{table:auctions_sp500} Results on closing auction trading of 250 randomly selected stocks belonging to the S\&P500. Reported variables are the transacted volume in the closing auction (in millions of dollars), the absolute value of the auction return (as defined in equation (\ref{eq:def_returns}), measured in bps) and the mean reversion coefficient $b$ resulting from the regression in equation (\ref{eq:regression}), when auctions with absolute returns above 10 bps and 30 bps are considered.
\\
\footnotesize{* The average over 2018 is significantly different from the average over 2017, $p<0.01$.}\\
\footnotesize{** The mean reversion coefficient $b$ is significant on the 0.05-level.} 
}
\end{table}
To check how closing prices behave for American stocks, we again estimate the model of equation (\ref{eq:regression}). In the third and fourth row of table \ref{table:auctions_sp500} the resulting mean reversion coefficients are shown. Clearly, the mean reversion coefficients are for American stocks in 2017 far more negative than for European stocks in 2017. This can have several reasons. Firstly, the closing auctions in the US are less important in terms of fraction of the total transacted volume (for S\&P500 stocks around 7\% of total transacted volume is transacted in the closing auction, while this is around $30\%$ for European stocks). Secondly, this can be explained by the differences in the auction mechanisms. As can be read in appendix \ref{appendix:closing auctions}, on NASDAQ/NYSE it is harder to act on new information than on Euronext, making closing prices possibly more efficient on Euronext (this reasoning is supported by the results of \cite{Comertonforderydge06}, who find that price formation is enhanced when more information is released). More important, the mean reversion coefficient for American stocks is in 2018 still significantly negative. It is thus observed that closing prices in the US are in 2018 still far from efficient, supporting the evidence that MiFID II triggered the observed effects in Europe.
\section{Conclusions} \label{sec:conclusions}
In this paper, effects of MiFID II on trading the constituents of the STOXX Europe 600 index are examined by comparing the years 2017 and 2018. The influence of the new tick size regime on liquidity, transacted volume, volatility and market stability is investigated. While prior research only focuses on intraday continuous trading, we also investigate tick size effects in the closing auction. It is shown that intraday liquidity is negatively (positively) affected by an increase (decrease) in tick size, except for very large orders. Interestingly, an increase in tick size improves liquidity in the closing auction, where there is no bid-ask spread, which rules out the adverse effects of an increase in tick size. Furthermore, an increase (decrease) in tick size leads to higher (lower) transacted volume, an effect that is in particular strong in the closing auction. Moreover, it is shown that an increase (decrease) in tick size has a positive (negative) effect on market stability, as it leads to less (more) microstructural noise. We also report significant effects of MiFID II that can not be attributed to the new tick size regime. Most remarkable, transacted volumes in the closing auction significantly increased since MiFID II, mainly driven by an increase in limit order volume. Along with this increase in volume, closing price deviations relative to the last minutes of continuous trading increased sharply. Remarkably, before MiFID II these closing price deviations consequently reverted overnight, but this overnight reversion completely vanished post-MiFID II, showing closing prices became more efficient. This implies that the extra volume in the closing auctions contributes to price formation and makes closing prices more efficient. These observations are not in line with an explanation based on the global trend of increasing passive investing, a reasoning that is supported by the fact that overnight reversion did not vanish in the US.  We did not find evidence that the extra volume was caused by the  introduced limits on dark pool trading or the new tick size regime, leaving open which exact part of MiFID II attracted volume to the closing auctions. We propose an explanation based on the best execution rule, which possibly forces market participants to trade in the closing auction, where every transaction is executed against a single price, making reporting obligations less cumbersome.  Our results show MiFID II has had significant effects on European stock markets (both expected and unexpected indirect consequences), that should be of interest to academics, policy makers and market practitioners.

%%%%%%%%%%%%%%%%%%%%%%%%%%%%%%%%%%%%%%%%%%%%%%%%%%%%%%%%%%%%%%%%%%%%%%%%%%%%%%

\newpage 
\begin{table}[h]
\footnotesize
\begin{tabular}{ p{3. cm} p{0.6 cm}p{.6 cm}p{1.2 cm} p{0.001cm} p{.6cm} p{.6cm} p{1.2cm} p{0.001cm} p{.6cm} p{.6cm} p{1.2 cm} }
\multicolumn{12}{c}{Statistics intraday liquidity and volume} \\
 \hline \\
 Variable  &\multicolumn{3}{c}{ts$\leftrightarrow$(control group)}& & \multicolumn{3}{c}{ ts$\uparrow$ }& &\multicolumn{3}{c}{ts$\downarrow$} \\
  \cline{2-4} \cline{6-8} \cline{10-12}\\
&2017 & 2018 & avg incr. && 2017 & 2018 & avg incr. &&2017 & 2018 & avg incr. \\
 \cline{2-4} \cline{6-8} \cline{10-12}\\
&\multicolumn{3}{c}{Large cap stocks}\\
 \cline{2-4} \cline{6-8} \cline{10-12}\\
Bid-Ask spread & 3.87 & 3.89 & 3.62\% && 4.83 & 6.75 & $52.66 \%^{*+}$ && 6.75 & 4.25 & $-35.96 \%^{*+}$ \\
Effective spread & 4.16 & 4.15 & 4.75\% && 4.18 & 6.31 & $48.74\%^{*+}$ && 7.07 & 4.39 & $-36.88\%^{*+}$ \\
Bid-Ask Vol1 & 0.277 & 0.181 & $-21.55\%$ && 0.0784 & 0.134 &$93.47\%^{*+}$ && 0.967 &0.253 & $-67.45\%^{*+}$ \\
Bid-Ask Vol3 & 0.9807 & 0.7160 & $-18.87\%$ && 0.2037 & 0.3753 & $138.3\%^{*+}$ && 1.657 & 0.5993 & $-64.55\%^{*+}$ \\
Bid-Ask Vol5 & 2.625 & 1.794 & $-23.20\%$ && 0.7308 & 1.220 &$118.5\%^{*+}$ && 7.435 & 2.300 & $-64.53\%^{*+}$ \\
Price impact 10k & 2.24 & 2.31 & $5.53\%$ && 3.39 & 4.37 & $38.40\%^{*+}$ && 3.51 & 2.39 & $-30.75\%^{*+}$\\
Price impact 20k & 2.59 & 2.73 & $7.26\%$ && 4.13 & 5.11 & $32.97\%^{*+}$ && 3.72 & 2.77 & $-25.37\%^{*+}$\\
Price impact 50k & 3.45 & 3.88 & $12.88\%$ && 6.02 & 7.01 & $30.16\%^{*+}$ && 4.35 & 3.78 & $-14.92\%^{*+}$\\
Price impact 100k & 4.66 & 5.52 & $18.36\%$ && 8.64 & 9.53 & $29.61\%^{*+}$ && 5.36 & 5.16 & $-6.79\%^{*+}$\\
Transacted volume & 88.61 & 94.35 & 7.43\% && 36.38 & 41.67 & $16.04\%^{*+}$ && 161.09 & 147.40 & $-0.492\%^{*+}$ \\
 &\multicolumn{3}{c}{Mid cap stocks}\\
 \cline{2-4} \cline{6-8} \cline{10-12}\\
Bid-Ask spread & 7.01 & 7.34 & 8.49\% && 6.88 & 9.32 & $43.93\%^{*+}$ && 8.30&6.06 & $-26.07\%^{*+}$ \\
Effective spread&6.44  & 7.54 & $16.93\%$ && 6.23& 8.37 & $40.51\%^{*+}$ && 8.10 & 5.98 & $-24.90\%^{*+}$ \\
Bid-Ask Vol1 &0.203 & 0.149 & $-16.96\%$ && 0.0469 & 0.0752 &$76.56\%^{*+}$ && 0.323 & 0.124 & $-60.14\%^{*+}$ \\
Bid-Ask Vol3 & 1.453 & 0.9534 & $-25.20\%$ && 0.3827 & 0.6454 & $122.5\%^{*+}$ && 4.396&1.222& $-68.08\%^{*+}$ \\
Bid-Ask Vol5 & 1.856 & 1.396 & $-15.88\%$ && 0.4072 & 0.7842 & $157.7\%^{*+}$ && 3.011 & 1.173 & $-60.91\%^{*+}$ \\
Price impact 10k & 4.32 & 4.57 & 10.71\% && 5.04 & 6.09 & $30.51\%^{*+}$ && 4.75 & 3.94 & $-16.20\%^{*+}$ \\
Price impact 20k & 5.27& 5.60 & 12.66\% && 6.62 & 7.65 & $27.36\%^{*+}$ && 5.47&4.93 & $-9.50\%^{*+}$ \\
Price impact 50k & 7.40 & 8.08 &16.32\% && 10.42 & 11.60 & $25.46\%^{+}$ && 7.43 & 7.36 & $-1.10\%^{*+}$ \\
Transacted volume & 39.56 & 45.73 & 18.32\% && 14.16 & 17.50 & 18.10\% && 58.38 & 52.54 & $8.89\%^{+}$ \\
 &\multicolumn{3}{c}{Small cap stocks}\\
 \cline{2-4} \cline{6-8} \cline{10-12}\\
 Bid-Ask spread & 8.77 & 9.43 & $10.60\%$ && 10.58 & 12.95 & $35.38\%^{*+}$ && 11.1 & 8.86 & $-14.02\%^{*+}$ \\
 Effective spread & 7.86 & 9.36 & $20.87\%$ && 9.46 & 12.75 & $43.33\%^{*+}$ && 10.13 & 8.43 & $-14.01\%^{*+}$\\
Bid-Ask Vol1 & 0.0840 & 0.0684 & $-10.76\%$ && 0.0532 & 0.0984 &$96.29\%^{*+}$ && 0.233 & 0.0919 & $-56.72\%^{*+}$ \\
Bid-Ask Vol3 & 0.4028 & 0.3101 & $-15.47\%$ && 0.2118 & 0.4506 & $191.9\%^{*+}$ && 1.095 & 0.3973 & $-63.96\%^{*+}$ \\
Bid-Ask Vol5 & 0.7954 & 0.6307 & $-13.11\%$ && 0.4446 & 0.9322 & $216.0\%^{*+}$ && 2.031 & 0.7970 & $-61.59\%^{*+}$\\
Price impact 10k & 6.34 & 6.69 & 13.18\% && 7.89 & 8.52 & $20.58\%^{+}$ && 6.74 & 6.58 & $1.23\%^{*+}$ \\
Price impact 20k & 8.29& 9.33 & 15.75\% && 10.60 & 10.87 & $16.50\%$ && 8.23& 8.69 & $6.26\%^{+}$ \\
Price impact 50k & 12.95 & 14.76 &17.85\% && 16.29 & 16.74 & $15.43\%$ && 11.98& 13.28 & $9.34\%$ \\
Transacted volume & 15.36 & 19.65 & $18.73\%$ && 9.73 & 13.94 & $35.46\%^{+}$ && 38.99 & 35.97 & $9.70\%$ \\
\hline
\end{tabular}
\caption{\label{table:liquidity} Descriptive statistics for the effects of tick size on intraday liquidity and transacted volume. Variables of interest are the quoted bid-ask spread and the effective bid-ask spread (both relative to the mid-price, measured in bps), the volume on the best bid and ask (Bid-Ask Vol1, measured in millions of euros), the volume on the first 3 and 5 levels of the order book (Bid-Ask Vol3/5, measured in millions of euros) and the virtual price impact of a market order of a given size (Price impact $x$, for $x=$10k, 20k, 50k and also 100k for large cap stocks, as defined in equation (\ref{eq:price_impact}), measured in basis points). Note that the last columns of the three blocks contain the average increase from 2017 to 2018 (\ie\ for every stock an increase from 2017 to 2018 is calculated and this is averaged over all stocks), not the increase in the reported averages over the stocks, see also the explanation in section \ref{sec:data}.\\
\footnotesize{* significantly different from the average increase in the control group `ts$\leftrightarrow$' of the same row($p<0.05$, t-test).\\
+ significantly different from the average increase of the control group `ts$\leftrightarrow$' of the same row($p<0.05$, Wilcoxon test).}}
\end{table}

\begin{table}[h]
\footnotesize
\begin{tabular}{ p{3 cm} p{0.001cm}  p{3.5cm} p{0.001cm} p{3.5cm} p{0.001cm} p{3cm} }

 \multicolumn{7}{c}{Regression results intraday liquidity and volume }\\
  \hline\\
Variable && $\beta_1$ (ts$\uparrow$, post-MiFID II) && $\beta_2$(ts$\downarrow$, post-MiFID II) && $\beta_5$ Post-MiFID II\\
\cline{3-3} \cline{5-5} \cline{7-7} \\
 &\multicolumn{2}{c}{Large cap stocks}\\
 \cline{3-3} \cline{5-5} \cline{7-7} \\
Bid-Ask spread && $2.251^{**}$ && $-2.589^{**}$ && $0.001663$ \\ 
Effective spread && $2.111^{**}$ && $-2.602^{**}$ && $-0.215^{*}$ \\
Bid-Ask Vol1 &&  $0.825^{**}$ && $-0.936^{**}$ && $ -0.243$ \\
Bid-Ask Vol3 && $1.011^{**}$ && $ -0.925^{**}$ && $-0.282^{**}$ \\
Bid-Ask Vol5 &&  $0.962^{**}$ && $-0.820^{**}$ && $ -0.221^{**}$\\
Price impact 10k && $0.911^{**}$ && $-1.181^{**}$ && $-0.00148$\\
Price impact 20k && $0.826^{**}$ && $-1.094^{**}$ && $0.0388^{**}$ \\
Price impact 50k && $0.36^{**}$ && $-0.986^{**}$ && $0.267^{**}$ \\
Price impact 100k && $-0.00180$ && $-1.00^{**}$ && $0.605^{**}$ \\
Transacted volume && $0.0470^{**}$ && $-0.0935^{**}$ && $0.0161^{**}$ \\
 &\multicolumn{2}{c}{Mid cap stocks}\\
 \cline{3-3} \cline{5-5} \cline{7-7} \\
 Bid-Ask spread && $2.113^{**}$ && $-2.769^{**}$ && $0.03715^{**}$\\
 Effective spread && $1.033^{**}$ && $-3.185^{**}$ && $0.675^{**}$ \\
 Bid-Ask Vol1 && $0.692^{**}$ && $ -0.772^{**}$ && $-0.171^{**}$ \\
 Bid-Ask Vol3 && $0.966^{**}$ && $ -0.858^{**}$ && $ -0.202^{**}$ \\
 Bid-Ask Vol5 && $0.974^{**}$ && $ -0.791^{**}$ && $-0.128^{**}$ \\
 Price impact 10k && $0.796^{**}$ && $-1.051^{**}$ && $-0.0249$\\
 Price impact 20k && $0.689^{**}$ && $-0.839^{**}$ && $0.00986$ \\
 Price impact 50k && $0.498^{**}$ && $-0.769^{**}$ && $0.232^{**}$ \\
 Transacted volume && -0.00588 && $-0.0753^{**}$ & &$0.0637^{**}$\\
  &\multicolumn{2}{c}{Small cap stocks}\\
 \cline{3-3} \cline{5-5} \cline{7-7} \\
 Bid-Ask spread&&  $2.360^{**}$ && $-3.122^{**}$ && $0.06397^{**}$ \\
 Effective spread&&$1.762^{**}$ && $-3.208^{**}$ && $0.988^{**}$ \\
 Bid-Ask Vol1 && $0.729^{**}$ && $ -0.749^{**}$ && $-0.127^{**}$ \\
 Bid-Ask Vol3 && $1.080^{**}$ && $-0.898^{**}$ && $-0.175^{**}$ \\
 Bid-Ask Vol5 && $1.084^{**}$ && $ -0.895^{**}$ &&$-0.0919^{**}$\\
 Price impact 10k && $0.188^{**}$ && $-0.825^{**}$ && $0.279^{**}$ \\
 Price impact 20k && $-0.581^{**}$ && $-0.612^{**}$ && $0.521^{**}$ \\
 Price impact 50k && $-1.072^{**}$ && $-0.570^{**}$ && $0.925^{**}$ \\
 Transacted volume && $0.132^{**}$ &&  $-0.0937^{**}$ && $0.0462^{**}$ \\
 \hline
\end{tabular}
\caption{\label{table:liquidity_regression} Results of the panel regression as specified in equation (\ref{eq:panel_regression}) for the following dependent variables: the quoted bid-ask spread and the effective spread (both relative to the mid-price, measured in bps), the volume on the best bid and ask (Bid-Ask Vol1, measured in euros, in log-scale), the volume on the first 3 and 5 levels of the order book (Bid-Ask Vol3/5, measured in euros, in log-scale) and the virtual price impact of a market order of a given size (Price impact $x$, for $x=$10k, 20k, 50k and also 100k for large cap stocks, as defined in equation (\ref{eq:price_impact}), measured in basis points).\\
\footnotesize{* the regression coefficient is significant on the $0.05$-level.}\\
\footnotesize{** the regression coefficient is significant on the $0.01$-level.}}
\end{table}

\begin{table}[h]
\footnotesize
\begin{tabular}{ p{3. cm} p{0.6 cm}p{.6 cm}p{1.2 cm} p{0.001cm} p{.6cm} p{.6cm} p{1.2cm} p{0.001cm} p{.6cm} p{.6cm} p{1.2 cm} }
\multicolumn{12}{c}{Statistics closing auction liquidity and volume} \\
 \hline \\
 Variable  &\multicolumn{3}{c}{ts$\leftrightarrow$(control group)}& & \multicolumn{3}{c}{ ts$\uparrow$ }& &\multicolumn{3}{c}{ts$\downarrow$} \\
  \cline{2-4} \cline{6-8} \cline{10-12}\\
&2017 & 2018 & avg incr. && 2017 & 2018 & avg incr. &&2017 & 2018 & avg incr. \\
 \cline{2-4} \cline{6-8} \cline{10-12}\\
&\multicolumn{3}{c}{Large cap stocks}\\
 \cline{2-4} \cline{6-8} \cline{10-12}\\
Post auction spread & 4.22 & 4.47 & 8.84\% && 5.20 & 5.84 & $14.41\%$ && 7.10 & 5.15 & $-23.29 \%^{*+}$ \\
Bid-Ask Vol1 & 1.271 & 1.259 & $3.03\%$ && 0.705 & 1.037 &$57.37\%^{*+}$ && 1.561 & 0.979 & $-67.45\%^{*+}$ \\
Bid-Ask Vol3 & 3.355 & 3.246 & $-2.06\%$ && 1.827  & 2.856 & $64.56\%^{*+}$ && 5.093 & 2.516 & $-46.16\%^{*+}$ \\
Bid-Ask Vol5 & 5.730 & 5.497 & $-2.71\%$ && 3.008 & 5.033 &$75.88\%^{*+}$ && 8.456 & 4.277 & $-43.64\%^{*+}$ \\
Illiquidity  & 0.531 & 0.491 & $-2.74\%$ && 3.074 & 1.861 & $-10.02\%^{*+}$ && 0.852 & 0.663 & $-7.62\%$ \\
Transacted volume & 26.70 & 33.03 & $25.41\%$ && 16.12 & 23.34 & $51.70\%^{+}$  && 19.51 & 23.38 & 29.07\% \\ 
 &\multicolumn{3}{c}{Mid cap stocks}\\
 \cline{2-4} \cline{6-8} \cline{10-12}\\
Post auction spread & 7.93 & 8.47 & 12.85\% && 7.52 & 8.27 & $8.98\%$ &&9.17 & 8.09 & $-9.14\%^{*+}$\\
Bid-Ask Vol1 &0.476 & 0.488 & $7.73\%$ && 0.305 & 0.489 &$77.89\%^{*+}$ && 0.697 & 0.436 & $-32.47\%^{*+}$ \\
Bid-Ask Vol3 & 1.392 & 1.344 & $-2.65\%$ && 0.803 & 1.362& $91.95\%^{*+}$ && 2.375 &1.205& $-44.47\%^{*+}$ \\
Bid-Ask Vol5 & 2.438 & 2.330 & 4.13\% && 1.311 & 2.362 & $104.22\%^{*+}$ && 4.203 & 2.089 & $-43.55\%^{*+}$ \\
Illiquidity & 3.352 & 2.586 & $-11.25\%$ && 6.751 & 3.977 & $-25.05\%^{*+}$ && 2.259 & 1.875 &$-7.08\%$  \\
Transacted volume & 5.86 & 7.48 &  35.04\% && 4.63 & 6.64 & $67.03\%^{*+}$ && 6.30 & 7.50 & $27.16\%^{+}$\\
 &\multicolumn{3}{c}{Small cap stocks}\\
 \cline{2-4} \cline{6-8} \cline{10-12}\\
 Post auction spread & 8.61 & 10.56 & $30.97\%$ && 8.88& 10.04 & $15.05\%$ && 9.36 & 9.67 & $5.13\%^{*+}$ \\
Bid-Ask Vol1 & 0.285 & 0.289 & $4.23\%$ && 0.176 & 0.280 &$72.42\%^{*+}$ && 0.428 & 0.266 & $-31.83\%^{*+}$ \\
Bid-Ask Vol3 & 0.819 & 0.786 & $-1.06\%$ && 0.479 & 0.843 & $95.07\%^{*+}$ && 1.366 & 0.698 & $-41.96\%^{*+}$ \\
Bid-Ask Vol5 & 1.404 & 1.345 & $0.689\%$ && 0.788& 1.478 & $109.62\%^{*+}$ && 2.478 & 1.177 & $-44.23\%^{*+}$\\ 
Illiquidity & 8.812 & 6.497 & $-8.73\%$ && 11.31 & 7.923& $-19.57\%^{*+}$ && 5.597 & 5.407 & $15.95\%^{*+}$ \\
Transacted volume & 2.89 & 3.73 & 41.83\% && 2.17 & 3.11 & 50.24\% && 3.43 & 3.56 & $16.01\%^+$\\
\hline
\end{tabular}
\caption{\label{table:liquidity_auc} Descriptive statistics for the effects of tick size on closing auction liquidity and transacted volume.  Variables of interest are the post auction bid-ask spread (relative to the closing price, measured in bps), the volume on the best bid and ask just after clearing (Bid-Ask Vol1, measured in millions of euros), the volume on the first 3 and 5 levels of the order book just after clearing (Bid-Ask Vol3/5, measured in millions of euros) and the illiquidity ratio (as defined in equation (\ref{eq:amihud}), in bps). Note that the last columns of the three blocks contain the average increase from 2017 to 2018 (\ie\ for every stock an increase from 2017 to 2018 is calculated and this is averaged over all stocks), not the increase in the reported averages over the stocks, see also the explanation in section \ref{sec:data}.\\
\footnotesize{* significantly different from the average increase in the control group `ts$\leftrightarrow$' of the same row($p<0.05$, t-test).\\
+ significantly different from the average increase of the control group `ts$\leftrightarrow$' of the same row($p<0.05$, Wilcoxon test).}}
\end{table}

\begin{table}[h]
\footnotesize
\begin{tabular}{ p{3 cm} p{0.001cm}  p{3.5cm} p{0.001cm} p{3.5cm} p{0.001cm} p{3cm} }

 \multicolumn{7}{c}{Regression results closing auction liquidity and volume }\\
  \hline\\
Variable && $\beta_1$ (ts$\uparrow$, post-MiFID II) && $\beta_2$(ts$\downarrow$, post-MiFID II) && $\beta_5$ Post-MiFID II\\
\cline{3-3} \cline{5-5} \cline{7-7} \\
 &\multicolumn{2}{c}{Large cap stocks}\\
 \cline{3-3} \cline{5-5} \cline{7-7} \\
Post auction spread && $0.451^{**}$ && $-2.133^{**}$ && $0.289^{**}$ \\
Bid-Ask Vol1 && $0.503^{**}$ && $-0.517^{**}$ && $-0.0667^{**}$ \\
Bid-Ask Vol3 && $0.539^{**}$ && $-0.675^{**}$ && $ -0.0620^{**}$\\
Bid-Ask Vol5 && $0.595^{**}$ && $-0.622^{**}$ &&$-0.0602^{**}$\\
Illiquidity && $-0.0551^{*}$ && $-0.0206$ && $-0.0790^{**}$\\
Transacted volume && $0.102^{**}$ && $-0.0729^{**}$ && $0.229^{**}$ \\
 &\multicolumn{2}{c}{Mid cap stocks}\\
 \cline{3-3} \cline{5-5} \cline{7-7} \\
 Post auction spread && $0.0981$ && $-1.642^{**}$ && $0.587^{**}$\\
Bid-Ask Vol1 && $0.561^{**}$ && $-0.564^{**}$ && $-0.0241^{**}$\\
Bid-Ask Vol3 && $0.594^{**}$ && $-0.666^{**}$ &&  $-0.0221^{**}$
\\
Bid-Ask Vol5 && $0.631^{**}$ && $ -0.650^{**}$ && $-0.0123^{**}$ \\
Illiquidity && $ -0.201^{**}$ && $ 0.0473$ && $-0.123^{**}$ \\
Transacted volume && $0.225^{**}$ && $ -0.0968^{**}$ && $0.263^{**}$ \\
  &\multicolumn{2}{c}{Small cap stocks}\\
 \cline{3-3} \cline{5-5} \cline{7-7} \\
 Post auction spread&&  $-0.974^{*}$ && $-1.455^{**}$ && $1.887^{**}$ \\
 Bid-Ask Vol1 &&  $0.593^{**}$ && $-0.403^{**}$ &&  $-0.0835^{**}$ \\
 Bid-Ask Vol3 && $0.659^{**}$ && $ -0.549^{**}$ && $-0.0657^{**}$\\
 Bid-Ask Vol5 && $ 0.700^{**}$ && $-0.624^{**}$ && $-0.0378^{**}$ \\
 Illiquidity && $-0.140^{**}$ && $0.280^{**}$ &&  $-0.107^{**}$ \\
 Transacted volume && $ 0.164^{***}$ && $-0.195^{**}$ && $0.309^{**}$ \\
 \hline
\end{tabular}
\caption{\label{table:liquidity_regression_auc}
 Results of the panel regression as specified in equation (\ref{eq:panel_regression}) for the following dependent variables: the post auction bid-ask spread (relative to the closing price, measured in bps), the volume on the best bid and ask just after clearing (Bid-Ask Vol1, measured in euros, in log-scale), the volume on the first 3 and 5 levels of the order book just after clearing (Bid-Ask Vol3/5, measured in euros, in log-scale) and the illiquidity ratio (as defined in equation (\ref{eq:amihud}), in log-scale).\\
\footnotesize{* the regression coefficient is significant on the $0.05$-level.}\\
\footnotesize{** the regression coefficient is significant on the $0.01$-level.}}
\end{table}
\begin{table}[h]
\footnotesize
%\begin{center}
\begin{tabular}{ p{2.85 cm} p{0.8 cm}p{.8 cm}p{1.2 cm} p{0.001cm} p{.8cm} p{.8cm} p{1.2cm} p{0.001cm} p{.8cm} p{.8cm} p{1.2 cm} }
 \multicolumn{12}{c}{Panel A: Statistics intraday stability and volatility}\\
 \hline \\
 Variable  &\multicolumn{3}{c}{ts$\leftrightarrow$(control group)}& & \multicolumn{3}{c}{ ts$\uparrow$ }& &\multicolumn{3}{c}{ts$\downarrow$} \\
 \cline{2-4} \cline{6-8} \cline{10-12}\\
&2017 & 2018 & avg incr. && 2017 & 2018 & avg incr. &&2017 & 2018 & avg incr. \\
 \cline{2-4} \cline{6-8} \cline{10-12}\\
&\multicolumn{3}{c}{Large cap stocks}\\
 \cline{2-4} \cline{6-8} \cline{10-12}\\
Volatility & 0.0109 & 0.0135 & 17.28\% && 0.0111 & 0.0138 & 22.42\% && 0.0116 & 0.0134 & $13.31\%^{*+}$ \\
Nr.\ bid-ask updates & 8603 & 11829 & 78.42\% && 7939 & 6386 & $-21.89\%^{*+}$ && 3293 & 10998 & $281.3\%^{*+}$ \\
Nr.\ trades & 8111 & 8546 & 5.36\% && 5539 & 5178 & $-7.48\%^{*+}$ &&
6570 & 7904 & $24.48\%^{*+}$ \\
Trade size & 13249 & 12816 & $-1.03\%$ && 8297 & 9880 & $26.46\%^{*+}$ && 25973 & 20241 & $-19.88\%^{*+}$ \\
 &\multicolumn{3}{c}{Mid cap stocks}\\
 \cline{2-4} \cline{6-8} \cline{10-12}\\
Volatility & 0.00998 & 0.0116 & 25.40\% && 0.00964 & 0.0117 & 24.87\% && 0.0110 & 0.0124 & $17.11\%^{*+}$ \\
Nr.\ bid-ask updates & 2730 & 4230 & 69.70\% && 4465 & 3619 & $-17.28\%^{*+}$ && 2043 & 5717 & $210.7\%^{*+}$\\
Nr.\ trades &  3094 & 3486 & 12.70\% && 2989 & 2585 & $-10.05\%^{*+}$ && 3307 & 4047 & $28.98\%^{*+}$ \\
Trade size & 12423 & 12404 & $3.33\%$ && 4929 & 6391 & $30.92\%^{*+}$ && 13642 & 11526&$-16.22\%^{*+}$ \\
 &\multicolumn{3}{c}{Small cap stocks}\\
 \cline{2-4} \cline{6-8} \cline{10-12}\\
 Volatility & 0.0122 & 0.0152 & 25.27\% && 0.0117 & 0.0148 & $27.68\%$ && 0.0138 & 0.0167 & $21.70\%$\\
Nr.\ bid-ask updates & 2582 & 3740 & 62.72\% && 2589 & 2109 & $-17.90\%^{*+}$ && 2107 & 5510 & $238.5\%^{*+}$ \\
Nr.\ trades & 2373 & 2447 & 8.89\% && 1640 & 1586 & $1.03\%$ && 2888& 3279 & $27.85\%^{*+}$ \\
Trade size & 6716 & 7190 & 7.05\% && 6363 & 8468 & $34.63\%^{*+}$ && 11326 & 9709 & $-15.41\%^{*+}$ \\
\\
 \multicolumn{12}{c}{Panel B: Statistics closing auction stability and volatility}\\
 \hline \\
 Variable  &\multicolumn{3}{c}{ts$\leftrightarrow$(control group)}& & \multicolumn{3}{c}{ ts$\uparrow$ }& &\multicolumn{3}{c}{ts$\downarrow$} \\
 \cline{2-4} \cline{6-8} \cline{10-12}\\
&2017 & 2018 & avg incr. && 2017 & 2018 & avg incr. &&2017 & 2018 & avg incr. \\
 \cline{2-4} \cline{6-8} \cline{10-12}\\
&\multicolumn{3}{c}{Large cap stocks}\\
 \cline{2-4} \cline{6-8} \cline{10-12}\\
Absolute auc return & 12.39 & 15.05 & $23.23\%$ && 12.91 & 15.93 & 24.19\% && 14.37 & 14.99 & $9.20\%^{*+}$ \\ 
Nr.\ ip updates &163& 198& 21.81\% && 108 &111& $3.11\%^{*+}$ && 118 &182 &$61.26\%^{*+}$ \\
&\multicolumn{3}{c}{Mid cap stocks}\\
 \cline{2-4} \cline{6-8} \cline{10-12}\\
Absolute auc return & 13.44 & 16.48 & $25.17\%$ && 13.96 & 16.98 & 22.77\% && 14.26 & 15.96 & $15.58\%^{*+}$ \\ 
Nr.\ ip updates& 84 & 89 & $8.15\%$ &&  66 & 64 & $0.037\%^{*+}$ &&  83 &106 & $32.35\%^{*+}$\\
&\multicolumn{3}{c}{Small cap stocks}\\
 \cline{2-4} \cline{6-8} \cline{10-12}\\
Absolute auc return & 14.09 & 17.58 & $28.75\%$ && 14.52 & 18.10 & 26.21\% && 14.39 & 17.78 & $25.30\%$ \\ 
Nr.\ ip updates  & 67 & 70 & $7.19\%$ &&  57 & 54 & $-3.63\%^{*+}$ && 72 & 80 & $11.93\%$ \\
\hline
\end{tabular}
%\end{center}
\caption{\label{table:volatility and market stability} Descriptive statistics for the effects of tick size on market stability and volatility. Variables of interest for intraday trading are daily volatility (measured as the daily sum of 1 minute mid price returns), the average daily number of best bid- or ask-price updates, the average daily number of trades and the average trade size (in euros). Variables of interest for the closing auction are the absolute auction return (the absolute value of the return in equation (\ref{eq:def_returns}), measured in bps) and the average number of indication price updates in the closing auction. Note that the last columns of the three blocks contain the average increase from 2017 to 2018 (\ie\ for every stock an increase from 2017 to 2018 is calculated and this is averaged over all stocks), not the increase in the reported averages over the stocks, see also the explanation in section \ref{sec:data}.\\
\footnotesize{* significantly different from the average increase in the control group `ts$\leftrightarrow$' of the same row ($p<0.05$, t-test).\\
+ significantly different from the average increase of the control group `ts$\leftrightarrow$' of the same row ($p<0.05$, Wilcoxon test).}}
\end{table}

\begin{table}[h]
\footnotesize
\begin{tabular}{ p{3 cm} p{0.001cm}  p{3.5cm} p{0.001cm} p{3.5cm} p{0.001cm} p{3cm} }
\multicolumn{7}{c}{Panel A: Regression results intraday stability and volatility}\\
  \hline\\
Variable && $\beta_1$ (ts$\uparrow$, post-MiFID II) && $\beta_2$(ts$\downarrow$, post-MiFID II) && $\beta_5$ Post-MiFID II\\
\cline{3-3} \cline{5-5} \cline{7-7} \\
 &\multicolumn{2}{c}{Large cap stocks}\\
 \cline{3-3} \cline{5-5} \cline{7-7} \\
Volatility && $0.000451^{**}$ && $-0.000344^{**}$ && $ 0.00159^{**}$ \\
Nr.\ bid-ask updates && $ -5774^{**}$ && $4432^{**}$ && $3176^{**}$ \\
Nr.\ trades && $-1078^{**}$ && $1279^{**}$ && $22.75$ \\
Trade size && $1621.73^{**}$ && $ -4982.52^{**}$ && $ -555.29^{**}$ \\
 &\multicolumn{2}{c}{Mid cap stocks}\\
 \cline{3-3} \cline{5-5} \cline{7-7} \\
Volatility && $3.56*10^{-5}$ && $-0.000797^{**}$ && $0.00259^{**}$ \\
Nr.\ bid-ask updates &&  $-2403^{**}$ && $2213^{**}$ && $1175^{**}$ \\
Nr.\ trades && $ -805^{**}$ && $505^{**}$ && $1.62$ \\
Trade size && $1505.04^{**}$ && $-1835.80^{**}$ && $-369.48^{*}$\\
  &\multicolumn{2}{c}{Small cap stocks}\\
 \cline{3-3} \cline{5-5} \cline{7-7} \\
Volatility && $ 0.000344^{**}$ && $4.92*10^{-5}$ && $0.00284^{**}$ \\
Nr.\ bid-ask updates && $ -1826^{**}$ && $2304^{**}$ && $900^{**}$ \\
Nr.\ trades && $ -396^{**}$ && $435^{**}$ &&$ -171$ \\
Trade size && $1118.27^{**}$ && $ -1887.02^{**}$ && $ -356.29^{**}$ \\
\\
\multicolumn{7}{c}{Panel B: Regression results closing auction stability and volatility}\\
  \hline\\
Variable && $\beta_1$ (ts$\uparrow$, post-MiFID II) && $\beta_2$(ts$\downarrow$, post-MiFID II) && $\beta_5$ Post-MiFID II\\
\cline{3-3} \cline{5-5} \cline{7-7} \\
 &\multicolumn{2}{c}{Large cap stocks}\\
 \cline{3-3} \cline{5-5} \cline{7-7} \\
 Absolute auc return && $0.571^{*}$ && $-2.091^{**}$ && $2.801^{**}$\\
 Nr.\ ip updates && $-37.51^{**}$ &&$29.33^{**}$ && $ 34.31^{**}$ \\
 &\multicolumn{2}{c}{Mid cap stocks}\\
 \cline{3-3} \cline{5-5} \cline{7-7} \\
 Absolute auc return  && $0.284$ && $-1.419^{**}$ && $3.192^{**}$ \\
 Nr.\ ip updates && $-7.64^{**}$ && $18.34^{**}$ && $4.72^{**}$ \\
  &\multicolumn{2}{c}{Small cap stocks}\\
 \cline{3-3} \cline{5-5} \cline{7-7} \\
 Absolute auc return && $0.276$ && $-0.0107$ && $3.473^{**}$ \\
 Nr.\ ip updates && $-7.29^{**}$ && $6.25^{**}$ && $2.19^{**}$ \\
 \hline
\end{tabular}
\caption{\label{table:stability_regression} Results of the panel regression as specified in equation (\ref{eq:panel_regression}) for the following dependent variables:  volatility (measured as the daily sum of 1 minute mid price returns), the average daily number of best bid- or ask-price updates, the average daily number of trades and the average trade size (in euros), the absolute auction return (the absolute value of the return in equation (\ref{eq:def_returns}), measured in bps) and the average number of indication price updates in the closing auction.\\
\footnotesize{* the regression coefficient is significant on the $0.05$-level.}\\
\footnotesize{** the regression coefficient is significant on the $0.01$-level.}}
\end{table}
\appendix

\section{Different Closing Auction Mechanisms}
\label{appendix:closing auctions}
Here we will discuss the mechanisms of the closing auctions in Europe and the US and their differences. As can be read below in more detail, the most important difference is the possibility of cancellations and modifications of pending orders based on new information, which is possible on Euronext, but not on NASDAQ, and on NYSE only through floor brokers. Not all constituents of the STOXX600 index are traded on Euronext, but closing auction mechanisms are very similar on the other important European exchanges (LSE, XETRA, SIX), so we will look at Euronext as an example.
\subsubsection*{Euronext\footnote{The rules of the Euronext auctions can be found in the harmonized rule book \citep[][section 4303]{EN19}.}}
On Euronext, the call phase of the closing auction starts at 5:30 pm, and lasts until a random time between 5:35:00 pm and 5:35:30 pm. During this call phase orders are accumulated without giving rise to transactions. Orders can be entered, modified and cancelled during this call phase, while during this whole period indicative price information is being published continuously. The indicative price at a given time is the price that would be the auction price if the auction stopped at that time, given the current order book. After the call phase has ended, the system will determine the closing price, which is determined as the price that maximizes the transacted volume. If there are several prices with equal transactable volume, the price that is closest to the last automated trade is taken. After the closing price is set and the order book is cleared (where all transactions are made against the closing price and market orders are ranked based on time-priority and have priority over limit orders, which are then ranked based on price-time priority), a phase of Trading-At-Last starts, which basically means that orders can be entered for execution at the closing price and for that price only. 

\subsubsection*{NASDAQ/NYSE\footnote{For the auction rules on NASDAQ, see \eg\ \url{https://www.nasdaqtrader.com/content/ProductsServices/Trading/Crosses/openclose_faqs.pdf}, for the auction rules on NYSE see \eg\ \url{https://www.nyse.com/publicdocs/nyse/markets/nyse/NYSE_Opening_and_Closing_Auctions_Fact_Sheet.pdf}}}
 On NASDAQ, orders for the closing auction can be entered the whole day, until the start of the auction at 3:55 pm. Two types of orders can be entered all day, market-on-close (MOC) orders and limit-on-close orders (LOC). MOC orders are executed against the closing price, whatever that price is, LOC orders are entered with a limit and are only executed if the closing price is equal or better than the limit price. At 3:55 pm NASDAQ starts to display the net order imbalance information. Between 3:55 pm and 3:58 pm it is not allowed to cancel or modify pending orders anymore, but participants can still enter LOC orders (as long as they are not more aggresive than the reference price at 3:55 PM, otherwise they are capped on this reference price). Then after 3:58, until the closing of the market at 4:00 pm, only imbalance orders (IO) can be entered, which are orders that can only be entered against the order imbalance. At 4:00 pm the closing price is determined by the system. The system determines the price that maximizes the transacted volume. If this does not give a single closing price, the system determines the price that minimizes the order imbalance. If this still does not lead to a single price, the system takes the price that is closest to the NASDAQ inside bid-ask midpoint. For execution, MOC orders are executed with time priority, then (if still possible) LOC orders are executed with price-time priority. The NYSE closing auction is very similar to the one on NASDAQ. However, on NYSE there are floor brokers that have priority, adding extra complexity to the system. The biggest difference with NASDAQ is the existence of so-called D-quotes, which can only be used by floor brokers and can be entered, modified and cancelled until 3:59:25 pm, regardless of their side of the order imbalance. Also, they are hidden from the imbalance information until 3:55 pm. At 4:00 pm, the closing price is determined in a similar way as on NASDAQ.
\section{Dark Pool Suspensions} \label{appendix:darkpools}
As already mentioned in the introduction of this paper, an important part of MiFID II is the DVC mechanism which allows only 4\% of a stock's yearly transacted volume to be transacted in a single dark pool, and only 8\% in all dark pools together. When these limits are violated for a particular stock, this stock gets suspended, meaning that this stock can not be traded in a dark pool anymore for a certain period of time. The European Securities and Markets Authority (ESMA) published the first suspensions on March 3, 2018. The list of suspended stocks can be used to test whether volume moved from dark pools to the regular exchanges for these suspended stocks. We split the stocks in the STOXX600 index in 2 groups: a group consisting of suspended stocks and a group consisting of stocks that are not suspended. The second group acts as a control group for the first group. In table \ref{table:darkpools} the change in transacted volume on the main exchanges from before March 3, 2018 to after this date is shown for the two groups, again separated for small, mid and large cap stocks. 
\begin{table}[h]
\footnotesize
\begin{center}
\begin{tabular}{ p{2.5cm} p{2.5 cm}p{2.5cm} p{2.5cm} }
 \multicolumn{4}{c}{Average increase from pre to post in transacted closing auction volume} \\
 \hline
 &\multicolumn{1}{c}{Small cap stocks} & \multicolumn{1}{c}{Mid cap stocks} &\multicolumn{1}{c}{Large cap stocks} \\
Suspended &  $31.03\%$  & $21.28\%$ & $23.82\%$ \\
 Not suspended &  $25.88\%$ &  $15.68\%$ & $34.89\%$ \\
 
  \multicolumn{4}{c}{Average increase from pre to post in transacted intraday volume} \\
 \hline
  &\multicolumn{1}{c}{Small cap stocks} & \multicolumn{1}{c}{Mid cap stocks} &\multicolumn{1}{c}{Large cap stocks} \\
Suspended & $14.47\%$ &  $11.61\%$ &$-1.579\%$ \\
 Not suspended & $17.18\%$ &   $4.227\%$ &  $4.525\%$ \\
 \hline
\end{tabular}
\end{center}
\caption{\label{table:darkpools} How the transacted volume in closing auctions and intraday changed after the first DVC-suspensions, for suspended and not suspended stocks. None of the increases for the suspended stocks is significantly higher than the increases for the not suspended stocks, all $p$-values$>0.15$.}
\end{table}
\begin{table}[h]
\footnotesize
\begin{center}
\begin{tabular}{ p{5 cm} p{2.5 cm} p{2.5 cm} p{2.5 cm} }
 \multicolumn{3}{c}{Results panel regression dark pool suspensions}\\
 \hline \\
Variable & Small cap stocks & Mid cap stocks & Large cap stocks \\
$\beta_6$ intraday volume & $0.0168^*$ & $0.0268^*$ &$-0.0188^{*}$ \\
$\beta_6$ closing auction volume & $0.0128^*$ & $0.0226^*$ & $-0.0351^*$ \\
\hline
\end{tabular}
\end{center}
\caption{\label{table:regression_darkpools} Results of the panel regression in equation (\ref{eq:panel_regression}), with the added dark pool suspension term in equation (\ref{eq:darkpoolterm}), for intraday transacted volume  and transacted volume in the closing auction (both in log-scale).
\\
\footnotesize{* The regression coefficient $\beta_6$ is significant on the 0.01-level.}\\
}
\end{table} 
If volume was moved from dark pools to the main exchanges after the suspensions, one would expect an increase in transacted volumes on the main exchanges for the suspended stocks, significantly higher than the increase in transacted volumes for the not suspended stocks. However, the results in table \ref{table:darkpools} suggest that this is not the case. Although volumes in the closing auction did increase significantly, this is also the case for not suspended stocks, and the increase is not significantly higher for suspended stocks (all $p$-values $>$ 0.15). The same holds intraday: for small and mid cap stocks, the intraday transacted volumes increased, but again not significantly more for suspended stocks. For large cap stocks the intraday transacted volumes even slightly decreased for suspended stocks. To complement this pairwise comparison, we also run again a panel regression, including stock fixed effects. To be more precise, we add a term 
\begin{align} \label{eq:darkpoolterm}
\beta_6 \mathbf{1}_S(i,t) 
\end{align}
 to the regression in equation (\ref{eq:panel_regression}), where $\mathbf{1}_S$ is an indicator that equals 1 if the stock $i$ is suspended on day $t$. We run the regression for transacted volume, intraday and in the closing auction. If suspension of dark pool trading caused volume to move to the main exchanges, we should find large positive and significant values for $\beta_6$. In table \ref{table:regression_darkpools} the results are shown, indicating that the dark pool suspensions did not cause significant volume to move to the main exchanges. Although $\beta_6$ is significantly positive for small and mid cap stocks, it is not economically significant. For instance, dark pool suspensions are only associated with an increase in closing auction volume in 2.26\% for mid caps. Furthermore, $\beta_6$ is even significantly negative for large cap stocks.
We conclude that the DVC regulations did not lead to the desired result: there is no evidence that significant volume did move from dark pools to the main exchanges after suspensions of stocks. This is in line with results by \cite{Johannetal19}, who also find that little volume from dark pools returned to the main exchanges after MiFID II. Instead, volumes moved to systems more similar to dark pools, such as periodic auction mechanisms and systematic internalisers\footnote{ The definition of which is given by MiFID II, Article 4, which states ``\emph{systematic internaliser means an investment firm which, on an organised, frequent systematic and substantial basis, deals on own account when executing client orders outside a regulated market, an MTF or an OTF without operating a multilateral system}
"}.
\end{document}